\documentclass[sigconf,nonacm]{acmart}

\renewcommand\footnotetextcopyrightpermission[1]{}
\usepackage{listings}
\usepackage{xspace}
\usepackage{wrapfig}
\usepackage{adjustbox}
\usepackage[normalem]{ulem}
\usepackage{algorithmic}
\usepackage[linesnumbered,ruled,vlined]{algorithm2e}
\usepackage{xcolor,color, colortbl}
\usepackage{tcolorbox}
\usepackage{subcaption}
\usepackage{pifont}

\usepackage{caption}
\newcommand{\distance}{7pt}
\AtBeginDocument{%
  \providecommand\BibTeX{{%
    \normalfont B\kern-0.5em{\scshape i\kern-0.25em b}\kern-0.8em\TeX}}}

\begin{document}


\newcommand{\overallAvgGain}{13X\xspace} 
\newcommand{\allBugsStudied}{74\%\xspace} 
\newcommand{\percentFailAnyway}{13\%\xspace} 
\newcommand{\percentHotSwapFails}{11\%\xspace} 
\newcommand{\percentDegenerate}{24\%\xspace} 
\newcommand{\percentTimeOutDelay}{17\%\xspace} 

\newcommand{\gv}{G\&V\xspace}

\newcommand{\lingming}[1]{\textcolor[rgb]{1.0,0.0,0.0}{#1}}
\newcommand{\ali}[1]{\textcolor[rgb]{0.0,0.0,1.0}{#1}}
\newcommand{\Comment}[1]{}
\newcommand{\omitlink}[1]{}

\newcommand{\codeIn}[1]{\texttt{#1}}
\newcommand{\uniapr}{UniAPR\xspace}
\newcommand{\apr}{APR\xspace}
\newcommand{\prapr}{PraPR\xspace}
\newcommand{\jaid}{JAID\xspace}
\newcommand{\skfix}{SketchFix\xspace}
\newcommand{\capgen}{CapGen\xspace}
\newcommand{\simfix}{SimFix\xspace}
\newcommand{\acs}{ACS\xspace}
\newcommand{\tbar}{TBar\xspace}
\newcommand{\jmutrepair}{jMutRepair\xspace}
\newcommand{\jgenprog}{jGenProg\xspace}


\newcommand{\arja}{Arja}
\newcommand{\kali}{Kali-A}
\newcommand{\genprog}{GenProg-A}
\newcommand{\rsrepair}{RSRepair-A}
\newcommand{\gprog}{GenProg}

\newcommand{\cardumen}{Cardumen}
\newcommand{\jkali}{jKali}

\newcommand{\dynamoth}{Dynamoth}
\newcommand{\nopol}{Nopol}

\newcommand{\avatar}{AVATAR}
\newcommand{\kpar}{kPar}
\newcommand{\fixminer}{FixMiner}
\newcommand{\prophet}{Prophet}

\newcommand{\Chart}{\textsf{Chart}\xspace}
\newcommand{\Math}{\textsf{Math}\xspace}
\newcommand{\Lang}{\textsf{Lang}\xspace}
\newcommand{\Time}{\textsf{Time}\xspace}
\newcommand{\Closure}{\textsf{Closure}\xspace}
\newcommand{\Mockito}{\textsf{Mockito}\xspace}


\SetKw{Continue}{continue}
\SetKw{Break}{break}

\lstset{escapeinside={*@}{@*},}
\lstset{language=Java,
        columns=fullflexible,
        basicstyle={\small\ttfamily},
        keywordstyle=\bfseries,
        breaklines=true,
        commentstyle=\color{gray},
        morecomment=[s][\color{javadocblue}]{/**}{*/},
}

\definecolor{GreenClover}{rgb}{0.61,0.78,0.62}
\definecolor{MistyRose}{rgb}{1.0,0.89,0.88}
\definecolor{GrayOne}{gray}{0.9}
\definecolor{GrayTwo}{gray}{0.8}
\definecolor{GrayThree}{gray}{0.7}
\definecolor{GrayFour}{gray}{0.6}
\definecolor{GrayFive}{gray}{0.5}


\newcommand{\prog}{\mathcal{P}}
\newcommand{\locs}{\mathcal{L}}
\newcommand{\progs}{\mathbb{P}}
\newcommand{\patch}{\mathcal{P'}}
\newcommand{\tests}{\mathcal{T}}
\newcommand{\test}{{t}}
\newcommand{\stat}{\mathcal{R}}
\newcommand{\progsl}{\progs_{left}}
\newcommand{\jvm}{\mathcal{JVM}}
\newcommand{\cname}{\mathcal{C}_{patched}}
\newcommand{\oldFile}{\mathcal{F}_{orig}}
\newcommand{\newFile}{\mathcal{F}_{patched}}
\newcommand{\status}{status}

\newcommand{\pat}{$\mathcal{P}$}
\newcommand{\clinit}{\codeIn{<clinit>()}}
\newcommand{\pclinit}{\codeIn{uniapr\_clinit()}}

\newcommand{\parabf}[1]{\noindent\textbf{#1}}
\newcommand{\hytt}[1]{\texttt{\hyphenchar\font=\defaulthyphenchar #1}}

\newcommand{\unsafeRatio}{5.41\%\xspace}
\newcommand{\resetOver}{10\%\xspace}

\title{Fast and Precise On-the-fly Patch Validation for All}


\author{Lingchao Chen}
\affiliation{%
  \institution{The University of Texas at Dallas}
}
\email{lingchao.chen@utdallas.edu}

\author{Lingming Zhang}
\affiliation{%
  \institution{The University of Texas at Dallas}
}
\email{lingming.zhang@utdallas.edu}

\begin{abstract}
Generate-and-validate (\gv) automated program repair (APR) techniques have been extensively studied during the past decade. Meanwhile, such\Comment{ \gv APR} techniques can be extremely time-consuming due to manipulation of the program code to fabricate a large number of patches and also repeated executions of tests on patches to identify potential fixes.
\prapr, a recent \gv APR technique, reduces these costs by modifying program code directly at the level of compiled bytecode, and further performing \emph{on-the-fly patching} by allowing multiple patches to be tested within the same JVM session. However, \prapr is limited due to its pattern-based, bytecode-level nature and it is basically unsound/imprecise as it assumes that patch executions do not change global JVM state and affect later patch executions on the same JVM session.\Comment{since patches involving large-scope changes can be hard to generate at the bytecode level}
Inspired by the \prapr work, we propose a unified patch validation framework, named \uniapr, which aims to speed up the patch validation for both bytecode and source-code \apr via on-the-fly patching; furthermore, \uniapr addresses the imprecise patch validation issue by resetting the JVM global state via runtime bytecode transformation.\Comment{which allows us to combine different APR techniques, thereby taking advantage of their strengths, and that reduces the cost of patch validation by avoiding unnecessary restart of JVM processes Our implementation is now publicly available on Maven Central.} We have implemented \uniapr as a\omitlink{ publicly available} fully automated Maven Plugin. We have also performed the first study of on-the-fly patch validation for state-of-the-art source-code-level APR.
\Comment{Inspired by the \prapr work, we perform the first study of on-the-fly patching for state-of-the-art APR work at the source-code level. Furthermore, we have addressed various challenges for applying on-the-fly patching to existing APR, such resetting the JVM status between test executions.}
Our experiments show the first empirical evidence that vanilla on-the-fly patch validation can be imprecise/unsound; in contrast, our \uniapr framework can speed up state-of-the-art \apr by over an order of magnitude without incurring any imprecision in patch validation, enabling all existing APR techniques to explore a larger search space to fix more bugs in the near future. Furthermore, \Comment{we also demonstrate that }\uniapr directly enables hybrid source and bytecode APR to fix substantially more bugs than all state-of-the-art APR techniques (under the same time limit) in the near future.
\end{abstract}





\maketitle

\section{Introduction}
Software bugs are inevitable in modern software systems, costing trillions of dollars in financial loss and affecting billions of people \cite{bib:Cambridge}. Meanwhile, software debugging can be extremely challenging and costly, consuming over half of the software development time and resources \cite{bib:WhitePaper}. Therefore, a large body of research efforts have been dedicated to automated debugging techniques \cite{bib:wong2016FLSurvey,monperrus2018automatic,gazzola2017automatic}. Among the existing debugging techniques,
automated program repair \cite{goues2019automated} (\apr) techniques hold the promise of reducing debugging effort by suggesting likely patches for buggy programs with minimal human intervention, and have been extensively studied in the recent decade. Please refer to the recent surveys on \apr for more details \cite{monperrus2018automatic,gazzola2017automatic}. 

Generate-and-validate (\gv) \apr refers to a practical category of APR techniques that attempt to fix the bugs by first generating a pool of patches and then validating the patches via certain rules and/or checks \cite{gazzola2017automatic}. A patch is said to be \textit{plausible} if it passes all the checks. Ideally, we would apply formal verification \cite{bib:nielson2019formal} techniques to guarantee correctness of generated patches. However, in practice, formal specifications are often unavailable for real-world projects, thus making formal verification infeasible. In contrast, testing is the prevalent, economic methodology of getting more confidence about the quality of software \cite{ammann2016introduction}. Therefore, the vast majority of recent \gv APR techniques leverage developer tests as the criteria for checking correctness of the generated patches~\cite{gazzola2017automatic}, i.e., \emph{test-based} \gv \apr.

Two main costs are associated with such test-based \gv APR techniques: (1) the cost of manipulating the program code to fabricate/generate a patch based on certain transformation rules; (2) repeated executions of all the developer tests to identify plausible patches for the bugs under fixing. Since the search space for APR is infinite and it is impossible to triage the elements of this search space due to theoretical limits, test-based \gv APR techniques usually lack clear guidance and act almost in a brute-force fashion: they usually generate a huge pool of patches to be validated and the larger the program the larger the set of patches to be generated and validated. This suggests that the speed of patch generation and validation plays a key role in scalability of the \apr techniques, which is one of the most important challenges in designing practical APR techniques \cite{phdxuan18}. Therefore, apart from introducing new, more effective, transformation rules, some APR techniques have been proposed to mitigate the aforementioned costs. For example, \jaid \cite{chen2017contract} uses mutation schema to fabricate meta-programs that bundle multiple patches in a single source file, while \skfix \cite{hua2018towards} uses sketches \cite{lezama2008program} to achieve a similar effect. However, such techniques mainly aim to speed up the patch generation time, while patch validation time has been shown to be dominant during \apr \cite{bib:mehne2018accelerating}.\Comment{; \capgen \cite{wen2018context} prioritizes patches based on certain heuristics so that patches that are more likely to be correct are tested first.} Most recently, \prapr \cite{ghanbari2019practical} aims to reduce both patch generation and validation time -- it reduces the cost of patch generation by modifying program code directly at the level of compiled JVM bytecode, and reduces the cost of patch validation by avoiding expensive process creation/initialization via reusing Java Virtual Machine (JVM) sessions across patches.\Comment{\prapr also reduces patch validation cost by reordering tests so that cheaper tests, as well as tests that are more likely to fail again, are executed first. This is in lines with test case reordering in mutation testing [?].}  \Comment{\lingming{don't mention test reordering stuff in this work since that is orthogonal to our main contribution and will confuse the reviewers, e.g., is our speedup due to on-the-fly patching or reordering?}}

We have empirical evidence that shows the optimizations offered by state-of-the-art APR tools make patch generation and validation much faster than before. However, compared to other techniques, the speed-up due to \prapr is huge (e.g. it is 90+X faster than \skfix). A careful analysis of the architecture of \prapr reveals that this remarkable speed-up is not just because of the way it generates patches\Comment{, or reorders test cases}, but is also because the tool is using the HotSwap technique \cite{oracleInstrAPI} to validate all the generated patches on-the-fly on the same JVM instead of creating a separate JVM process for each patch. This turns out to be a dominant factor. Process creation overhead in systems like JVM is especially pronounced as virtually all the optimization tasks, as well as Just-In-Time (JIT) compilation \cite{aho2006compilers}, are done at runtime. Therefore, compared to natively executed programs, JVM-based programs are expected to take relatively longer time to warm up. Furthermore, during \apr, a similar set of used bytecode files are loaded, linked, and initialized again and again for each patch. This suggests repeated executions of patches in separate processes (which is the dominant approach in mainstream APR techniques) for JVM-based based languages waste a significant amount of time that could be otherwise spent on applying more sophisticated patch generation rules or exploring more of the search space. 

\prapr, although is effective, suffers from two major problems: (1) it is not flexible due to its bytecode-level nature; and (2) it is unsafe as it might report unsound/imprecise patch validation results.\Comment{It was easier for me to describe it here as I am going to motivate the reader\lingming{you can only mention (1) here, and we can show (2) when we talk about the experimental results since it is our study's contribution to find such case that HotSwap may bring side effects}} The first problem is best illustrated by inserting a factor in an arithmetic expression. For example, mutating \codeIn{a*b+(a-b)} to \codeIn{a*b+(a*c-b)} turns out to be a non-trivial program analysis task at the bytecode level. This is because Java compiler might reorder the bytecode instructions based on the priority of arithmetic operators or avoid repeated memory accesses when the variables \codeIn{a} and \codeIn{b} are declared \codeIn{final} by doing simple optimization of loading the variables once and duplicating their values on the JVM stack. Either of these optimizations would make the such a program transformation hard to implement efficiently as locating the right place for inserting the instructions corresponding to the second multiplication is hard.\Comment{Another challenging situation arises when dealing with common repair task of inserting \codeIn{break} and \codeIn{continue} statements in nested loops and switch statements inside loops.} Apparently this task would be trivial if we had modified the program at the level of source code by manipulating Abstract Syntax Trees (ASTs) of the programs. Also, it has been widely recognized as notoriously challenging to perform large-scale changes at the bytecode level, making the set of bugs fixable by \prapr rather limited. The second problem is that the global JVM state may be \emph{polluted} by earlier patch executions, making later patch execution results unreliable. For example, some patches may modify some static fields, which are used by some later patches sharing the same JVM. Note that although the original \prapr{} work does not have such imprecision issue due to the specific limited types of patches supported (by bytecode \apr), we do find instances of such imprecision in our study (detailed shown in Section~\ref{sec:prec}).

\Comment{\prapr might also report patches as plausible fixes that evidently do not pass the all the test cases. This situation arises when the state of patch validation process is infected by the side-effects of executing some patch. Regardless of the size of the change that the APR technique makes to the program, this situation tends to happen in programs that lack a proper mechanism for setup/teardown before and after test case that might have side-effects or might be dependent on side-effects. Please note that we do not encounter this situation when we test the program using build systems like Maven \cite{maven} or Ant \cite{ant}. This is because, in these systems, although we run all the test cases in a single process, we do it only once and the effects of running test cases do not propagate to the subsequent runs.}

Motivated by the strengths and weaknesses of \prapr, in this paper, we propose a unified test-based patch validation framework, named \uniapr, that reduces the cost of patch validation by avoiding unnecessary restarts of JVM for all existing bytecode or sourcode-level \apr techniques. \uniapr achieves this by using a single JVM for patch validation, as much as possible, and depends on JVM's dynamic class redefinition feature (a.k.a. the HotSwap mechanism and Java Agent technology) to only reload the patched bytecode classes on-the-fly for each patch. Furthermore, it also addresses the imprecision problem of \prapr by isolating patch executions via resetting JVM states after each patch execution via runtime bytecode transformation. In this way, \uniapr not only substantially speeds up all state-of-the-art \apr techniques at the source-code level (enabling them to explore larger search space to fix more bugs in the near future), but also provides a natural framework to enable \emph{hybrid \apr{}} to combine the strengths of various \apr techniques at both the source-code and bytecode levels.

\uniapr has been implemented as a\omitlink{ publicly available} fully automated Maven~\cite{maven} plugin\omitlink{~\cite{uniapr}}, to which almost all existing state-of-the-art Java APR tools can be attached in the form of patch generation \textit{add-ons}. We have constructed add-ons for representative APR tools from different \apr families. Specifically, we have constructed add-ons for \capgen~\cite{wen2018context}, \simfix \cite{jiang2018Shaping}, and \acs \cite{xiong2017precise} that are modern representatives of template-/pattern-based \cite{goues2012genprog,debroy2010Using,le2016history}, heuristic-based \cite{barr2014plastic}, and constraint-based \cite{xuan2017nopol,nguyen2013semfix} techniques. With these techniques, we have conducted the first extensive study of on-the-fly validation of patches generated at the source code level. Our experiments show that \uniapr can speed up state-of-the-art \apr systems (i.e., \capgen, \simfix, and \acs) by over an order of magnitude without incurring any imprecision in patch validation, enabling all existing APR techniques to explore a larger search space to fix more bugs in the near future.

We envision a future wherein all existing \apr tools (like \simfix~\cite{jiang2018Shaping}, \capgen~\cite{wen2018context}, and \acs~\cite{xiong2017precise}) and major \apr frameworks (like ASTOR \cite{martinez2016ASTOR} and Repairnator \cite{repairnator2019monperrus}) are leveraging this framework for patch validation\Comment{and new techniques are built on top of this}. In this way, researchers will need only to focus on devising more effective algorithms for better exploring the patch search space, rather than spending time on developing their own components for patch validation, as we can have a unified, generic, and much faster framework for all. Furthermore, our \uniapr directly enables hybrid source and bytecode \apr to fix substantially more bugs than all state-of-the-art \apr techniques (under the same time limit) in the near future.

In summary, this paper makes the following contributions:
\begin{itemize}
    \item \textbf{Framework.} We introduce the first unified on-the-fly patch validation framework, \uniapr, to speed up \apr techniques for JVM-based languages at both the source and bytecode levels.  
    \item \textbf{Technique.} We show the first empirical evidence that on-the-fly patch validation can be imprecise/unsound, and introduce a new technique to reset the JVM state right after each patch execution to address such issue.
    \Comment{that is applicable on an assortment of JVM-based APR techniques. We discuss and address the challenges constructing such a framework.}
    \item \textbf{Implementation.} We have implemented on-the-fly patch validation based on the JVM HotSwap mechanism and Java Agent technology \cite{oracleInstrAPI}, and implemented the JVM-reset technique based on the ASM bytecode manipulation framework \cite{owASM}; the overall \uniapr tool has been implemented as a practical Maven plugin\omitlink{ publicly available online~\cite{uniapr}}, and can accept different \apr{} techniques as patch generation add-ons to reduce their patch validation cost\Comment{ by relying on JVM's dynamic class redefinition to validate patches, instead of creating isolated processes for each patch}.
    \item \textbf{Study.} We conduct a large-scale study of the effectiveness of \uniapr on its interaction with state-of-the-art \apr tools from three different \apr families, demonstrating that \uniapr can speed up state-of-the-art \apr by over an order of magnitude (with precise validation results), and can enable hybrid \apr to directly combine the strengths of different \apr tools.
\end{itemize}

The rest of this paper is organized as follows. We introduce the necessary background information in Section \ref{sec:preliminaries}. In Section \ref{sec:approach}, we introduce the details of the proposed \uniapr technique. Next, we present our experimental setup and result analysis in Section \ref{sec:setup} and Section~\ref{sec:res}. Finally we discuss related work in Section \ref{sec:related} before we conclude the paper in Section \ref{sec:conclude}.

\section{Background}\label{sec:preliminaries}
In this section, we set the scene by introducing some background for better understanding this work. More specifically, we first talk about the current status of automated program repair (Section~\ref{sec:aprPrelim}); then, we talk about Java Agent and HotSwap, on which our \uniapr{} work is built on (Section~\ref{sec:jvmPrelim}).

\subsection{Automatic Program Repair}\label{sec:aprPrelim}
Automatic program repair (APR) aims to suggest likely patches for buggy programs to reduce the manual effort during debugging. Based on the actions taken for fixing a bug, state-of-the-art APR techniques can be divided into: (1) techniques that monitor the execution of a system to find deviations from certain specifications, and then heal the system by modifying its runtime state in case of any abnormal behavior \cite{perkins2009automatically,long2014automatic}; (2) generate-and-validate (\gv) techniques that attempt to fix the bug by first generating a pool of patches and validating the patches via certain rules and/or checks \cite{goues2012genprog,nguyen2013semfix,wen2018context,ghanbari2019practical,jiang2018Shaping,martinez2016ASTOR}. Generated patches that can pass all the tests/checks are called \emph{plausible} patches. However, not all plausible patches are the patches that the developers want. Therefore, these plausible patches are further manually checked by the developers to find the final \emph{correct} patches (i.e., the patches semantically equivalent to developer patches). Among these, \gv techniques, especially those based on tests, have gained popularity as testing is the dominant way for detecting bugs in practice, while very few real-world systems are based on rigorous and up-to-date formal specifications.

In recent years, a large number of APR-related research papers have been published in different software engineering and programming languages conferences and journals \cite{gazzola2017automatic,monperrus2018automatic}. These papers either introduce new techniques for generating high quality patches or study different aspects of already introduced techniques. Recently, Ghanbari et al. \cite{ghanbari2019practical} showed for the first time that the sheer speed of patch generation and validation gives an otherwise simplistic template-/pattern-based technique like \prapr a discrete advantage, in that it allows the tool to explore more of the entire search space in an affordable amount of time. The explored search space is more likely to contain plausible and hence correct patches.

The speed of patch generation and validation in \prapr comes from on-the-fly patch generation which is possible due to two features: (1) bytecode-level patch generation; (2) on-the-fly patch validation based on dynamic class redefinition. \prapr generates patches by directly modifying programs at the level of compiled JVM bytecode \cite{lindholm2017JVM}. This allows the tool to bypass expensive tasks of parsing and modifying ASTs, as well as type checking and compilation (which itself is preceded by several undoubtedly expensive accesses to secondary memory). Furthermore, the tool also avoids starting a new JVM session for validating each and every one of the patches. Instead, it creates a single JVM process and leverages the Java Agent technology and HotSwap mechanism offered by JVM to reload only the patched bytecode files for each patch without restarting the JVM. In this way, \prapr{} not only can avoid reloading (also including linking and initializing) all used classes for each patch (i.e., only the patched bytecode file(s) needs to be reloaded for each patch), but also can avoid the unnecessary JVM warm-up time (i.e., the accumulated JVM profiling information cross patches enables more and more code to be JIT-optimized and the already JIT-optimized code can also be shared across patches). In this paper, we further generalize the \prapr on-the-fly patching to all existing \apr systems (whereas prior work only applied it for bytecode \apr~\cite{ghanbari2019practical}), and also address the unsoundness/imprecision issues for such optimization. \Comment{dynamic class redefinition feature of JVM offered by Java instrumentation API \cite{oracleInstrAPI} to alter the behavior of already loaded classes. Java instrumentation API returns the old bytecode of the loaded class before updating their definition.} 

\subsection{Java Agent and HotSwap}\label{sec:jvmPrelim}
A Java Agent \cite{oracleInstrAPI} is a compiled Java program (in the form of a JAR file) that runs alongside of the JVM in order to intercept applications running on the JVM and modify their bytecode. Java Agent utilizes the instrumentation API \cite{oracleInstrAPI} provided by Java Development Kit (JDK) to modify existing bytecode that is loaded in the JVM. In general, developers can both (1) \emph{statically} load a Java Agent using \codeIn{-javaagent} parameter at JVM startup, and (2) \emph{dynamically} load a Java Agent into an existing running JVM using the Java Attach API. For example, to load it statically, the manifest of the JAR file containing Java Agent must contain a field \texttt{Premain-Class} to specify the name of the class defining \codeIn{premain} method. Such a class is usually referred to as an \codeIn{Agent class}. Agent class is loaded before any class in the application class is loaded and the \codeIn{premain} method is called before the main method of the application class is invoked.\Comment{\lingming{Java Agent also supports other ways, e.g., via agentmain, and I believe we also use agentmain since premain only modify systems before they are launched} \ali{The underlying HotSwapAgent of PIT uses that. We just use the parameter \texttt{Instrumentation} that is passed by the JVM}}
The method \texttt{premain} usually has the following signature:
\begin{center}
    \texttt{public static void premain(String agentArgs, Instrumentation inst)}
\end{center}
The second parameter is an object of type \texttt{Instrumentation} created by the JVM that allows the Java Agent to analyze or modify the\Comment{ definition of} classes loaded by the JVM (or those that are already loaded) before executing them. Specifically, the method \texttt{redefineClasses} of \texttt{Instrumentation}, given a \textit{class definition} (which is essentially a class name paired with its ``new'' bytecode content), even enables dynamically updating the definition of the specified class, i.e., directly replacing certain bytecode file(s) with the new one(s) during JVM runtime. This is typically denoted as the JVM HotSwap mechanism.
It is worth mentioning that almost all modern implementations of JVM (especially, so-called HotSpot JVMs) have these features implemented in them.

By obtaining \texttt{Instrumentation} object, we have a powerful tool using which we can implement a HotSwap Agent. As the name suggests, HotSwap Agent is a Java Agent and is intended to be executed alongside the patch validation process to dynamically reload patched bytecode file(s) for each patch. In order to test a generated patch during \apr, we can pass the patched bytecode file(s) of the patch to the agent, which \emph{swaps} it with the original bytecode file(s) of the corresponding class(es). Then, we can continue to run tests which results in executing the patched class(es), i.e., validating the corresponding patch. Note that subsequent requests to HotSwap Agent for later patch executions on the same JVM are always preceded by replacing previously patched class(es) with its original version.  In this way, we can validate all patches (no matter generated by source-code or bytecode \apr) on-the-fly sharing the same JVM for much faster patch validation\Comment{ (as demonstrated in the prior \prapr work~\cite{ghanbari2019practical})}.
\Comment{Specifically, to be consistent with Algorithm \ref{alg:prapr}, we may call our agent class \texttt{HotSwapAgent}. In this class, after obtaining the \texttt{Instrumentation} object in its \texttt{premain} method, we can store it in a field of the class. We can implement the method \texttt{replace} used in Algorithm \ref{alg:prapr} by simply loading the current definition the class to be replaced, redefining the class using the method \texttt{redefineClasses}, and returning the old definition.}

\begin{figure*}[t!]
    \centering
    \includegraphics[scale=0.4]{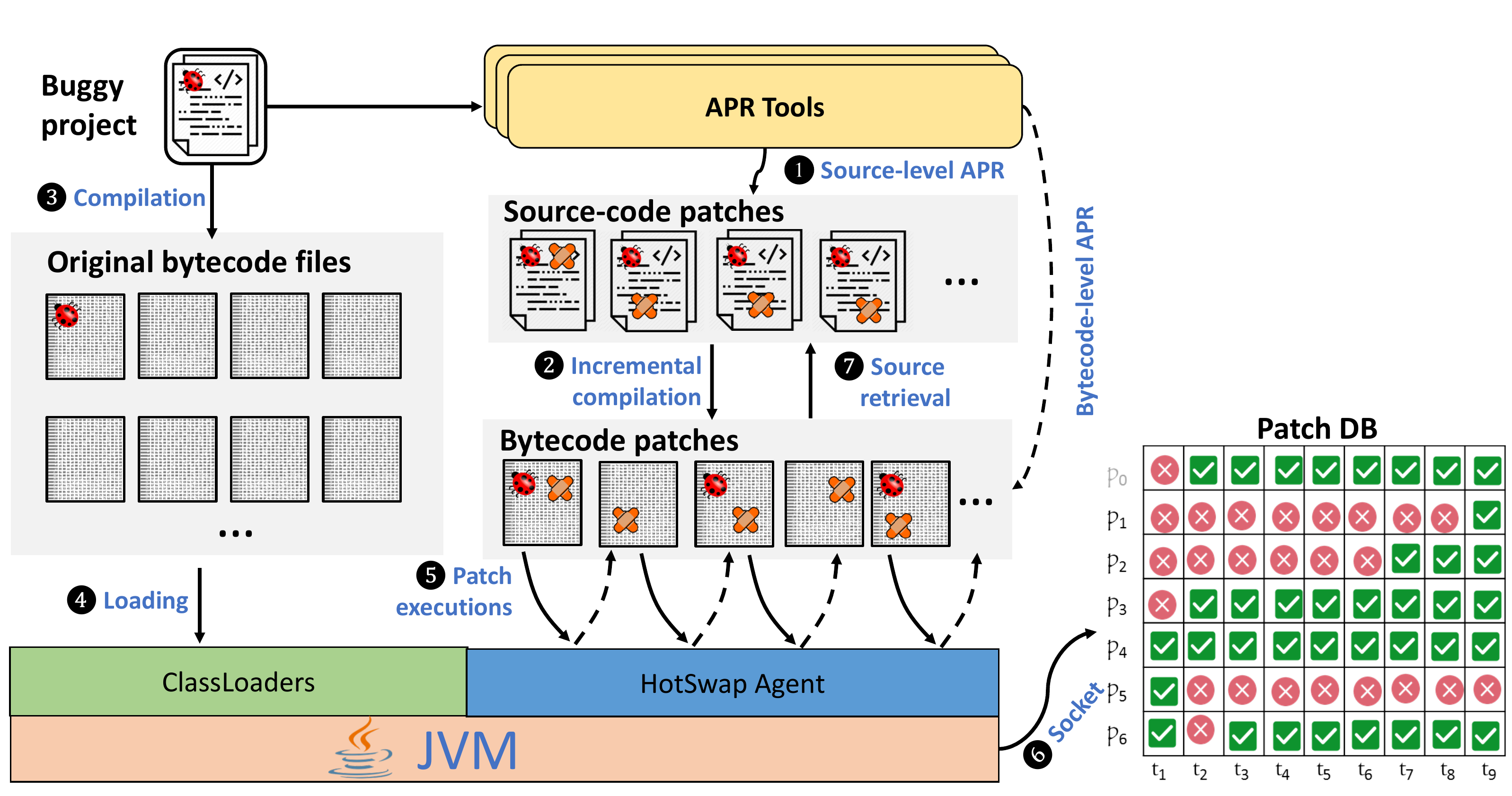}
    \caption{\Comment{Architectural view of \uniapr\lingming{you can make it more detailed to show how source-code patches from different tools can be transformed into bytecode patches and how those bytecode patches can be hotswapped via sharing the same JVM (and resetting the JVM for each patch)}}An architectural overview of \uniapr and its workflow}
    \label{fig:uniaprArch}
\end{figure*}
\section{Approach}\label{sec:approach}
In this section, we first talk about the overall approach for our \uniapr system (Section~\ref{sec:overview}). Then, we will talk about our detailed design for \emph{fast} patch validation via on-the-fly patching (Section~\ref{sec:otf}) as well as \emph{precise} patch validation via JVM reset (Section~\ref{sec:reset}).
\subsection{Overview}
\label{sec:overview}

Figure \ref{fig:uniaprArch} depicts an the overall flow of our \uniapr framework.
According to the figure, given a buggy project, \uniapr first leverages any of the existing \apr tools (integrated as \uniapr add-ons) to generate source-code level patches (marked with \ding{182}). Then, \uniapr performs incremental compilation to compile the patched source file(s) by each patch into bytecode file(s) (marked with \ding{183}). Note that, \uniapr is a unified framework and can also directly take the bytecode patches generated by the \prapr (and future) bytecode \apr technique (marked with the dashed line directly connecting \apr tools into bytecode patches). In this way, \uniapr has a pool of bytecode patches for patch validation. 

During the actual patch validation, \uniapr first compiles the entire buggy project into bytecode files (i.e., \codeIn{.class} files), and then loads all the bytecode files into the JVM through JVM class loaders (marked with \ding{184} and \ding{185} in the figure). Note that these two steps are exactly the same as executing the original tests for the buggy project. Since all the bytecode files for the original project are loaded within the JVM, for validating each patch, \uniapr only reloads the patched bytecode file(s) by that particular patch via the Java Agent technology and HotSpot mechanism, marked with \ding{186} (as the other unpatched bytecode files are already within the JVM). Then, the test driver can be triggered to execute the tests to validate against the patch without restarting a new JVM. After all tests are done for this patch execution, \uniapr will replace the patched bytecode file(s) with the original one(s) to revert to the original version. Furthermore, \uniapr also resets the global JVM states to prepare a clean JVM environment for the next patch execution (marked with the short dashed lines). The same process is repeated for each patch. Finally, the patch validation results will be stored into the patch execution database via socket connections  (marked with \ding{187}). Note that for any plausible patch that can pass all  the tests, \uniapr will directly retrieve the original source-level patch for manual inspection (marked with \ding{188}) in case the patch was generated by source-level \apr.

\Comment{applies a patch generation add-on on the file and generates a set of transformed source files. In order to validate the generated patches \uniapr runs all the test cases against each patch, but before that it needs to obtain \texttt{.class} files for the patches which is done by (incrementally) compiling each of the generated patches. The main novelty of this work is in the way it runs test cases against the patches. Vanilla testing approach, which is prevalent in state-of-the-art APR techniques, creates a separate process for each patch while existing HotSwap-based approaches run each test case oblivious of the fact that the test cases might have side-effects affecting the execution of one another. In fact this is the source of unsoundness in patch validation system of state-of-the-art \prapr. Therefore, in this work we \emph{reset} JVM status immediately after executing each test case. Please note that we do not rely on \emph{setup} and \emph{teardown} methods associated with test classes and test methods to clean up system state. This is because such methods are not intended to reset every possible state-preserving component of JVM as they are not supposed to be executed over and over again in the same process. To address that \uniapr explicitly resets system states by resetting static fields to they default values and invoking class initializers, resetting system properties, resetting JMX beans, etc. }

\Comment{which is in the form of a Maven plugin capable of interacting with different \textit{patch generation add-ons}. According to the figure, given a buggy source file, \uniapr applies a patch generation add-on on the file and generates a set of transformed source files. In order to validate the generated patches \uniapr runs all the test cases against each patch, but before that it needs to obtain \texttt{.class} files for the patches which is done by (incrementally) compiling each of the generated patches. The main novelty of this work is in the way it runs test cases against the patches. Vanilla testing approach, which is prevalent in state-of-the-art APR techniques, creates a separate process for each patch while existing HotSwap-based approaches run each test case oblivious of the fact that the test cases might have side-effects affecting the execution of one another. In fact this is the source of unsoundness in patch validation system of state-of-the-art \prapr. Therefore, in this work we \emph{reset} JVM status immediately after executing each test case. Please note that we do not rely on \emph{setup} and \emph{teardown} methods associated with test classes and test methods to clean up system state. This is because such methods are not intended to reset every possible state-preserving component of JVM as they are not supposed to be executed over and over again in the same process. To address that \uniapr explicitly resets system states by resetting static fields to they default values and invoking class initializers, resetting system properties, resetting JMX beans, etc. \ali{please explain your implementation here.}}
\Comment{
as a Maven plugin that interacts with different patch generation add-ons. The plugin, as part of the target program's POM file, receives a set of user configurations that instructs the tool how it should perform patch generation and validation. In particular, user-specified configuration includes the name of the patch generation add-on as well as other parameter like time-out threshold, whether or not the tool should double-check plausible patches in case it finds any, and whether or not the tool should fall back to vanilla patch validation in case it fails to find any plausible patches. We stress that patch generation strategy does not have to produced compiled bytecode for the patches. In fact, the POM file may have an an option on whether the patches produced by the patch generation add-on are already in JVM bytecode form or they need compilation (the compiler can also be specified). With that said, \uniapr is not limited to Java program.}

We have already constructed add-ons for three different APR tool representing three different families of APR techniques. These add-ons include \capgen~\cite{wen2018context} (representing pattern/template-based APR techniques), \simfix~\cite{jiang2018Shaping} (representing heuristic-based techniques), and \acs~\cite{xiong2017precise} (representing constraint-based techniques)\Comment{, and \tbar (representing pattern based techniques)}. Users of \uniapr can easily build a new patch generation add-on by implementing the interface \texttt{PatchGenerationPlugin} provided by the framework. For already implemented APR tools, this can be easily done by changing their source code so that the tools abandon validation of patches after generating and compiling them. \Comment{In case the source code of the APR tool is not available (like \capgen), the users can still make a \emph{wrapper program} that overrides the default behavior of the tool so that it does not run test cases against the generated patches anymore.}

\subsection{Fast Patch Validation via On-the-fly Patching}\label{sec:otf}
Algorithm~\ref{alg:uniapr} is a simplified description of the steps that \emph{vanilla} \uniapr (without JVM-reset) takes in order to validate candidates patches on-the-fly. The algorithm takes as inputs the original buggy program $\prog$, its test suite $\tests$, and the set of candidate patches $\progs$ generated by any \apr technique. The output is a map, $\stat$, that maps each patch into its corresponding execution result. The overall \uniapr algorithm is rather simple. \uniapr first initializes all patch execution results as unknown (Line 2). Then, \uniapr gets into the loop body and obtains the set of patches still with unknown execution results (Line 4). If there is no such patches, the algorithm simply returns since all the patches have been validated. Otherwise, it means this is the first iteration or the earlier JVM process gets terminated abnormally (e.g., due to timeout or JVM crash). In either case, \uniapr will create a new JVM process (Line 7) and start to evaluate the remaining patches in this new JVM (Line 8). 

We next talk about the detailed \codeIn{validate} function, which takes the remaining patches, the original test suite, and a new JVM as input. For each remaining patch $\patch$, the function first obtains the patched class name(s) $\cname$ and patched bytecode file(s) $\newFile$ within $\patch$ (Lines 11 and 12). \Comment{\lingming{Ali, how does your implementation load the original bytecode files before all patch execution? We should add that here} \ali{I will add details, but FYI for now, I am not explicitly loading them from disk. I put the target directory into the classpatch of the patch validator process and rely on JVM classloaders to load the classes on demand as the tests get executed.}}\Comment{It relies on JVM's default class loaders to load original bytecode files from the classpath.} Then, the function leverages our HotSwap Agent to replace the bytecode file(s) under the same class name(s) as $\cname$ with the patched bytecode file(s) $\newFile$; it also stores the replaced bytecode file(s) as $\oldFile$ to recover it later (Line 13). Note that our implementation will explicitly load the corresponding class(es) to patch (e.g., via \codeIn{Class.forName()}) if they are not yet available before swapping. In this way, the function can now execute the tests within this JVM to validate the current patch since the patched bytecode file(s) has already been loaded (Lines 14-26). If the execution for a test finishes normally, its status will be marked as \codeIn{Plausible} or \codeIn{Non-Plausible} (Lines 16-19); otherwise, the status will be marked as \codeIn{Error}, e.g., due to timeout or JVM crash (Lines 20-21). Then, $\patch$'s status will be updated in $\stat$ (Line 22). If the current status is \codeIn{Non-Plausible}, the function will abort the remaining test executions for the current patch since it has been falsified, and move on to the next patch (Line 24); if the current status is \codeIn{Error}, the function will return to the main algorithm (Line 26), which will restart the JVM. When the validation for the current patch finishes without the \codeIn{Error} status, the function will also recover the patched bytecode file(s) into the original one(s) to facilitate the next patch validation (Line 27).

\Comment{The algorithm is intended to be executed in a process as long as possible---i.e., until all the patches are validated or something uncontrollable, such as a JVM crash, or timeout happens. The implicit assumption in this algorithm is that each patch targets exactly one class. The algorithm takes each patch $q$, obtains the name the corresponding class (Line 4). It then dynamically updates the definition of the class so that the patched class gets loaded by the JVM (Line 5), runs the test cases and reports validation status (Lines 6-12) before restoring the old definition of the class (Line 14). Please note that between each test execution, the algorithm cleans up the state of JVM by invoking \texttt{doJVMReset} at Line 13.\ali{Please describe your implementation of this procedure.}\Comment{\lingming{the algorithm and detailed description shall be moved to a later UniAPR approach section; for here, just talk about PraPR's basic design and idea in text}}}

\begin{algorithm}[t!]
\scriptsize
    \SetKwProg{BeginFunc}{function}{:}{end}
    \SetKwProg{try}{try}{:}{}
    \SetKwProg{catch}{catch}{:}{end}
    \caption{\label{alg:uniapr} Vanilla on-the-fly patch validation in \uniapr}
    \KwIn{Original buggy program $\prog$, test suite $\tests$, and set of candidate patches $\progs$}
    \KwOut{Validation status $\stat:\progs\rightarrow\{\mathtt{PLAUSIBLE},\mathtt{NON-PLAUSIBLE},\mathtt{ERROR}\}$}
    \Begin{
        $\stat\leftarrow \progs\times \{\mathtt{UNKNOWN}\}$ \tcp*[l]{initialize the result function} 
        \While{$\mathtt{True}$} {
            $\progsl\leftarrow\{\patch\mid \patch\in\progs\wedge \stat(\patch)=\mathtt{UNKNOWN}\}$\tcp{get all the left patches not yet validated} 
            \If{$\progsl=\emptyset$}{
                \Return{$\stat$} \tcp{return if there is no left patches}
            }
            $\jvm\leftarrow\mathtt{createJVMProcess()}$\tcp{create a new JVM}
            $\mathtt{validate}(\progsl, \tests, \jvm))$ \tcp{validate the left patches on the new JVM}
        }
    }
    \BeginFunc{$\mathtt{validate}(\progsl, \tests, \jvm)$}{
        \For{$\patch$ in $\progsl$}{
            $\cname\leftarrow \mathtt{patchedClassNames(\patch)}$ \leavevmode \\
            $\newFile\leftarrow \mathtt{patchedBytecodeFiles(\patch)}$ \leavevmode \\
            $\oldFile\leftarrow\mathtt{HotSwapAgent.swap(}\jvm, \mathtt{\cname}, \newFile)$ \tcp{Swap in the patched bytecode files} 
            \For{$t$ in $\tests$}{
                \try{}{
                \uIf{\texttt{run(}$\jvm, t$\texttt{)} = \texttt{FAILING}}{
                    $\status\leftarrow\mathtt{NON-PLAUSIBLE}$
                }
                \uElse{ $\status\leftarrow\mathtt{PLAUSIBLE}$}
                }
                \catch{TimeOutException, MemoryError}{
                    $\status\leftarrow\mathtt{ERROR}$ 
                }
                $\stat\leftarrow \stat\cup\{\patch\rightarrow \status\}$ \leavevmode \\
                \If{$\status$ = \texttt{NON-PLAUSIBLE}}{
                    \Break \tcp{continue with the next patch when current one is falsified}
                }
                \If{$\status$ = \texttt{ERROR}}{
                    \Return{} \tcp{restart a new JVM when this current one timed out or crashed}
                }
            } 
            $\mathtt{HotSwapAgent.swap(}\jvm, \cname, \oldFile\mathtt{)}$\tcp{Swap back the original bytecode files}
        } 
    } 
\end{algorithm}

\Comment{\lingming{let's elaborate this more, for example, you haven't talked about the ReStart conditions (e.g., timeout, crash etc.).} The most important tools used in this algorithm are the subroutine \texttt{replace} of HotSwap agent and \texttt{doJVMReset}.\Comment{ We are going to give a tutorial introduction on how we can make a HotSwap agent in the following subsection.} The internals of such procedures is described in Section ?? and Section ??.}
\Comment{Having a patch generation strategy $G$ at hand, \uniapr proceed as follows.\lingming{let's make it an formal algorithm here, and describe that in natural language; the current format looks more like a lab report}
\begin{enumerate}
    \item Apply the add-on $G$ on the buggy program $P$ to get a set $Q$ of patches (in the form of compiled bytecode paired with the name of the patched classes), and supplemental information about the transformations done by $G$.
    \item If the patches in $Q$ are not already in JVM bytecode form, use the prescribed compiler to incrementally compile all the patches in $Q$ to get the corresponding set $Q_1$ of compiled patches.
    \item Update the status of all patches in $Q_1$ as \texttt{UNKNOWN}.
    \item Repeat until each patch $q\in Q_1$ is marked either as \texttt{PLAUSIBLE} or \texttt{NON-PLAUSIBLE}.
    \begin{enumerate}
        \item Let $Q'$ be the set of all patches $q\in Q_1$ such that $q$ is marked as \texttt{UNKNOWN}.
        \item Create a new validation process to execute Algorithm \ref{alg:prapr} on $P$, the set of test cases of $P$, and $Q'$ to get a mapping $s$ describing validation status of the patches in $Q'$.
        \item Wait until the process crashes or dies gracefully.
    \end{enumerate}
    \item Report patch validation status $s$, together with supplemental information to explain more about the patches that are marked as \texttt{PLAUSIBLE} in $s$.
\end{enumerate}
}
\subsection{Precise Patch Validation via JVM Reset}
\label{sec:reset}

\begin{figure}
\scriptsize
\center
\includegraphics[scale=0.4]{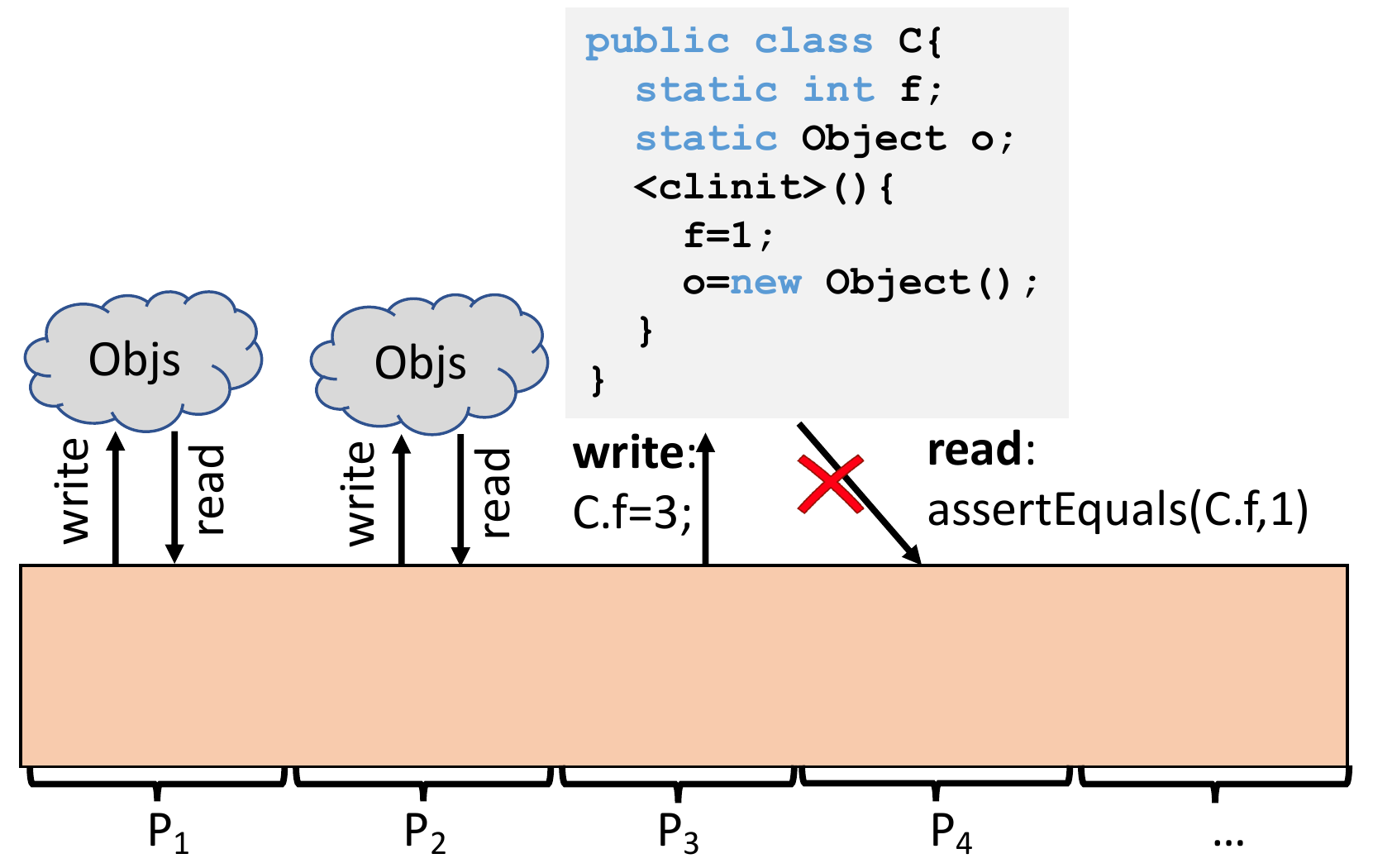}
\caption{\label{fig:vanilla}Imprecision under vanilla on-the-fly patch validation}
\end{figure}

\subsubsection{Limitations for vanilla on-the-fly patch validation}
The vanilla on-the-fly patch validation presented in Section~\ref{sec:otf} works for most patches of most buggy projects. The basic process can be illustrated via Figure~\ref{fig:vanilla}. In the figure, each patch (e.g., from \pat$_1$ to \pat$_4$) gets executed sequentially on the same JVM. It would be okay if every patch accesses and modifies the objects created by itself, e.g., \pat$_1$ and \pat$_2$ will not affect each other and the vanilla on-the-fly patch validation results for \pat$_1$ and \pat$_2$ will be the same as the ground-truth patch validation results. However, it will be problematic if one patch writes to some global space (e.g., static fields) and later on some other patch(es) reads from that global space. In this way, earlier patch executions will affect later patch executions, and we call such global space \emph{pollution sites}. To illustrate, in Figure~\ref{fig:vanilla}, \pat$_3$ write to some static field \codeIn{C.f}, which is later on accessed by \pat$_4$. Due to the existence of such pollution site, the execution results for \pat$_4$ will no longer be precise, e.g., its assertion will now fail since \codeIn{C.f} is no longer 1, although it may be a correct patch.

\subsubsection{Technical challenges}
We observe that accesses to static class fields are the main reason leading to imprecise on-the-fly patch validation. Ideally, we only need to reset the values for the static fields that may serve as pollution sites right after each patch execution. In this way, we can always have a clean JVM state to perform patch execution without restarting the JVM for each patch. However, it turns out to be a rather challenging task: 

\begin{figure}
\center
\begin{lstlisting}[language=JAVA, basicstyle=\ttfamily\scriptsize]
// org.joda.time.TestYearMonthDay_Constructors.java
public class TestYearMonthDay_Constructors extends TestCase {
    private static final DateTimeZone PARIS = DateTimeZone.forID("Europe/Paris");
    private static final DateTimeZone LONDON = DateTimeZone.forID("Europe/London");
    private static final Chronology GREGORIAN_PARIS =
            GregorianChronology.getInstance(PARIS);
    ...
\end{lstlisting}
\caption{\label{fig:field} Static field dependency}
\end{figure}
First, we cannot simply reset the static fields that can serve as pollution sites. The reason is that some static fields are \codeIn{final} and cannot be reset directly. Furthermore, static fields may also be data-dependent on each other; thus, we have to carefully maintain their original ordering, since otherwise the program semantics may be changed. For example, shown in Figure~\ref{fig:field}, \codeIn{final} field \codeIn{GREGORIAN\_PARIS} is data-dependent on another \codeIn{final} field, \codeIn{PARIS} under the same class within project Joda-Time \cite{jodaTime} from the widely studied Defects4J dataset \cite{bib:just2014defects4j}. The easiest way to keep such ordering and reset \codeIn{final} fields is to simply re-invoke the original class initializer for the enclosing class. However, according to the JVM specification, only JVM can invoke such static class initializers.
\begin{figure}
\center
\begin{lstlisting}[language=JAVA, basicstyle=\ttfamily\scriptsize]
// org.joda.time.TestDateTime_Basics.java
public class TestDateTime_Basics extends TestCase {
    private static final ISOChronology ISO_UTC = ISOChronology.getInstanceUTC();
    ...
// org.joda.time.chrono.ISOChronology.java
public final class ISOChronology extends AssembledChronology {
    private static final ISOChronology[] cFastCache;
    static {
        cFastCache = new ISOChronology[FAST_CACHE_SIZE];       
        INSTANCE_UTC = new ISOChronology(GregorianChronology.getInstanceUTC());
        cCache.put(DateTimeZone.UTC, INSTANCE_UTC);
    }
    ...
\end{lstlisting}
\caption{\label{fig:clinit} Static initializer dependency}
\end{figure}

Second, simply invoking the class initializers for all classes with pollution sites may not work. A naive way to reset the pollution sites is to simply trace the classes with pollution sites executed during each patch execution; then, we can simply force JVM to invoke the class initializers for all those classes after each patch execution. However, it can bring side effects in practice because the class initializers may also depend on each other. For example, shown in Figure~\ref{fig:clinit}, within Joda-Time, the static initializer of class \codeIn{TestDateTime\_Basics} depends on the static initializer of \codeIn{ISOChronology}. If \codeIn{TestDateTime\_Basics} is reinitialized earlier than \codeIn{ISOChronology}, then field \codeIn{ISO\_UTC} will no longer be matched with the newest \codeIn{ISOChronology} state. Therefore, we have to reinitialize all such classes following their original ordering if they had been 
executed on a new JVM.

Third, based on the above analysis, we basically have two choices to implement such system: (1) customizing the underlying JVM implementation, and (2) simulating the JVM customizations at the application level. Although it would be easier to directly customize the underlying JVM implementation, the system implementation will not be applicable for other stock JVM implementations. That said, we are only left with way of simulating the JVM customizations at the application level. 

\begin{figure}
\center
\includegraphics[scale=0.25]{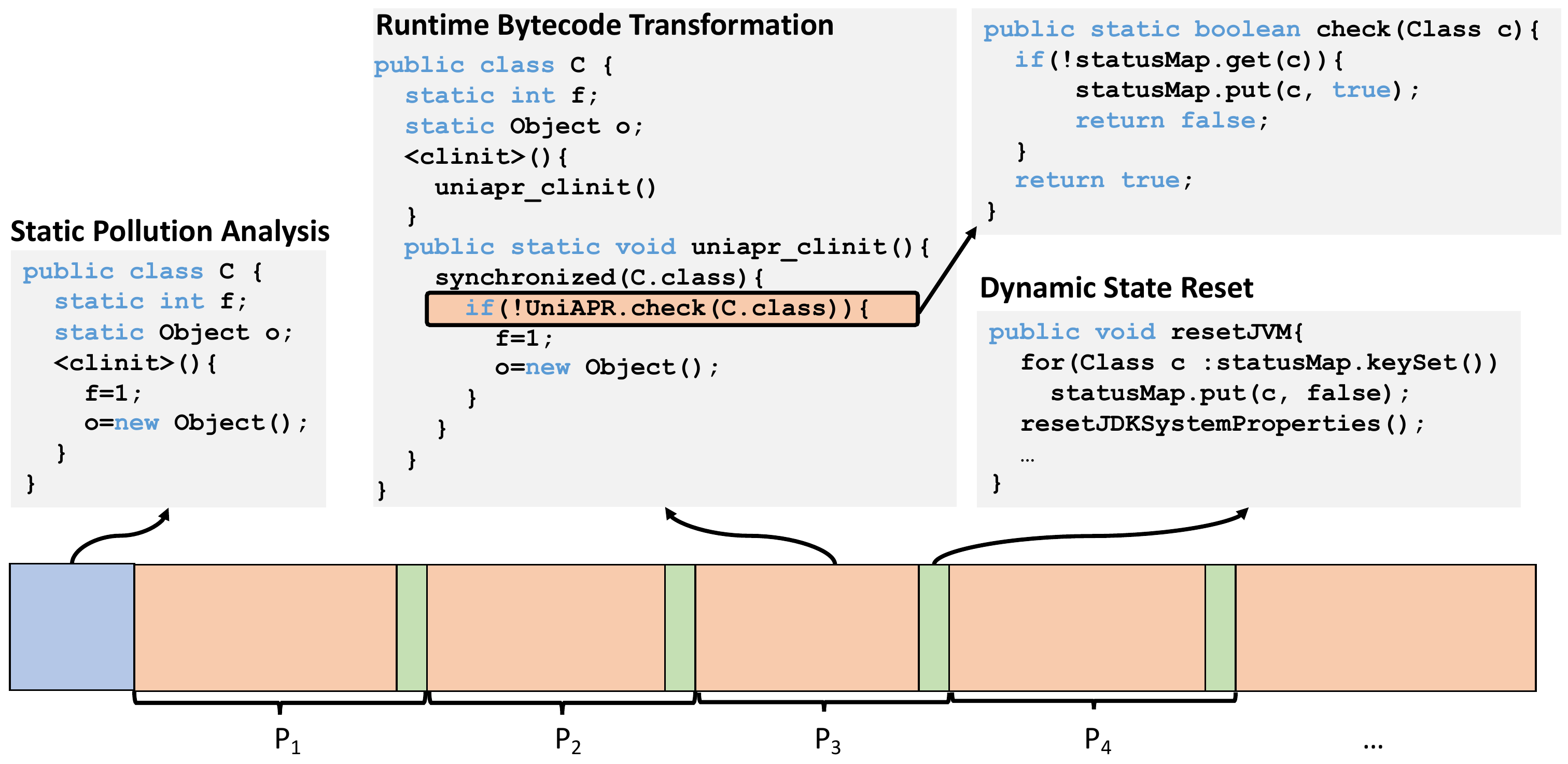}
\caption{On-the-fly patch validation via JVM reset}
\label{fig:reset}\end{figure}

\subsubsection{JVM reset via bytecode transformation} 
We now present our detailed approach for resetting JVM at the the application level. Inspired by prior work on speeding up traditional regression testing~\cite{bell2014unit}, we\Comment{our basic insight is to} perform runtime bytecode transformation to simulate JVM class initializations. The overall approach is illustrated in Figure~\ref{fig:reset}. We next present the detailed three phases as follows.

\parabf{Static Pollution Analysis.} Before all the patch executions, our approach performs lightweight static analysis to identify all the pollution sites within the bytecode files of all classes for the project under repair, including all the application code and 3rd-party library code. Note that we do not have to analyze the JDK library code since JDK usually provides public APIs to reset the pollution sites within the JDK, e.g., \codeIn{System.setProperties(null)} can be used to reset any prior system properties and \codeIn{System.setSecurityManager(null)} can be leveraged to reset prior security manager. The analysis basically returns all classes with non-\codeIn{final} \codeIn{static} fields or \codeIn{final} \codeIn{static} fields with non-primitive types (their actual object states in the heap can be changed although their actual references cannot be changed), since the states for all such static fields can be changed across patches. Shown in Figure~\ref{fig:reset}, the blue block denotes our static analysis, and class \codeIn{C} is identified since it has static fields \codeIn{f} and \codeIn{o} that can be mutated during patch execution.\Comment{Furthermore, our static analysis also trace the class-inheritance hierarchy among all possible classes.}

\begin{table}[h]
\scriptsize
    \centering
    \begin{tabular}{c|l}
    \hline
C1&\codeIn{T} is a class and an instance of \codeIn{T} is created\\
C2&\codeIn{T} is a class and a static method declared by \codeIn{T} is invoked.\\
C3&A static field declared by \codeIn{T} is assigned\\
C4&A static field declared by \codeIn{T} is used and the field is not a constant variable \\
C5&\codeIn{T} is a top level class, and an assert statement lexically nested within \codeIn{T} is executed\\
\hline
    \end{tabular}
    \caption{Class initialization conditions}
    \label{tab:clinit}
\end{table}

\parabf{Runtime Bytecode Transformation.} According to Java Language Specification (JSL)~\cite{javaspec}, static class initializers get invoked when any of the five conditions shown in Table~\ref{tab:clinit} gets satisfied. Therefore, the ideal way to reinitialize the classes with pollution sites is to simply follow the JSL design.\Comment{will talk about class hierarchy impact on reclinit} To this end, we perform runtime bytecode transformation to add class initializations right before any instance that falls in to the five conditions shown in Table~\ref{tab:clinit}. Note that our actual implementation also handles the non-conventional Reflection-based accesses to all potential pollution sites within classes. Since JVM does not allow class initialization at the application level, 
following prior work~\cite{bell2014unit} on speeding up normal test executions during regression testing, we rename the original class initializers (i.e., \codeIn{<clinit>()}) to be invoked into another customizable name (say \pclinit{}). Meanwhile, we still keep the original \codeIn{<clinit>()} initializers since JVM needs that for the initial invocation; however, now \codeIn{<clinit>()} initializers do not need to have any content except an invocation to the new \pclinit{}. Note that we also remove potential \codeIn{final} modifiers for pollution sites during bytecode transformation to enable reinitializations of \codeIn{final} non-primitive static fields. Since this is done at the bytecode level after compilation, the original compiler will still ensure that such \codeIn{final} fields cannot be changed during the actual compilation phase.

Now, we will be able to reinitialize classes via invoking the corresponding \pclinit{} methods. However, JVM only initializes the same class once within the same JVM, while now \pclinit{} will be executed for each instance satisfying the five conditions in Table~\ref{tab:clinit}. Therefore, we need to add the dynamic check to ensure that each class only get (re)initialized once for each patch execution. Shown in Figure~\ref{fig:reset}, the pink blocks denote the different patch executions. During each patch execution, the classes with pollution sites will be transformed at runtime, e.g., class \codeIn{C} will be transformed into the code block connected with the \pat$_3$ patch execution. Note that the pink code block denotes the dynamic check to ensure that \codeIn{C} is only initialized once for each patch\footnote{Note that this pseudo code is just for illustration, and our actual implementation manipulates arrays for faster and safe tracking/check.}. The pseudo code for the dynamic check is shown in the top-left of the figure. We can see that the check maintains a concurrent \codeIn{HashMap} for the classes with pollution sites and their status (\codeIn{true} means the corresponding class has been reinitialized). The entire initialization is also synchronized based on the \codeIn{class} object to handle concurrent accesses to class initializers; in fact, JVM also leverages a similar mechanism to avoid class reinitializations due to concurrency (despite implementing that at a different level). In this way, when the first request for initializing class \codeIn{C} arrives, all the other requests will be blocked. If the class has not been initialized, then only the current access will get the return value of \codeIn{false} to reinitialize \codeIn{C}, while all other other requests will get the return value of \codeIn{true} and skip the static class initialization. Furthermore, the static class initializers get invoked following the same order as if they were invoked in a new JVM.

\parabf{Dynamic State Reset.} After each patch execution, our approach will reset the state for the classes within the status \codeIn{HashMap}. In this way, during the next patch execution, all the used classes within the \codeIn{HashMap} will be reinitialized (following the check in Figure~\ref{fig:reset}). Note that besides the application and 3rd-party classes, the JDK classes themselves may also have pollution sites. Luckily, JDK provides such common APIs to reset such pollution sites without the actual bytecode transformation. In this way, our implementation also invokes such APIs to reset potential JDK pollution sites. Please also note that our system provides a public interface for the users to customize the reset content for different projects under repair. For example, some projects may require preparing specific external resources for each patch execution, which can be easily added to our public interface. In Figure~\ref{fig:reset}, the green strips denote the dynamic state reset, and the example reset code connected to \pat$_3$ simply resets the status for all classes within the status map as \codeIn{false} and also resets potential JDK pollution sites within classes.

\section{Experimental Setup}\label{sec:setup}
\subsection{Research Questions}\label{sec:rqs}
To thoroughly evaluate our \uniapr framework, in this study, we aim to investigate the following research questions:
\begin{itemize}
\item RQ1: How does vanilla on-the-fly patch validation perform for automated program repair?

\Comment{For this RQ, we can have two parts: efficiency analysis and effectiveness analysis. 
For efficiency analysis, let's just show the average patch number, original patch generation time, original patch execution time, new vanilla execution time, and average speedup (e.g., 5X) across all bugs for each tool. 
For effectiveness analysis, let's draw a small table to show the number of bugs with inconsistent results with the original APR tools. Let's also show some example code/bug to explain why on-the-fly does not have the same results with the original tools, this can help further motivate our study for JVM-reset.
}

\item RQ2: How does on-the-fly patch validation with jvm-reset perform for automated program repair? 

\Comment{
For this RQ, we also have efficiency and effectiveness analysis.
For efficiency analysis, we can show the average patch number, original execution time, vanilla PIT-style execution time with its average speedup, JVM-reset execuiton time with its average speedup, and JVM-restart execution time with its average speedup across all bugs for each tool. The goal is to show that JVM-reset, although slightly slower than the PIT-stype one, it can be much faster that JVM-restart.
For effectiveness analysis, we can also the number of bugs with inconsistent results with the original APR tools (I assume it should close to 0). Then, we can explain why JVM-reset works.
}

\end{itemize}
For both RQs, we study both the \emph{effectiveness} of \uniapr in reducing the patch validation cost, and the \emph{precision} of \uniapr in producing precise patch validation results.
\subsection{Benchmarks}
\begin{table}
\scriptsize
  \centering\small
  \begin{tabular}{|l||l|rrr|}
    \hline
    Sub. & Name & \#Bugs&\#Tests&LoC \\
    \hline\hline
    \Chart & JFreeChart & 26 &2,205&96K\\
    \Time & Joda-Time & 27 &4,130&28K\\
    \Lang & Apache commons-lang & 65 &2,245&22K\\
    \Math & Apache commons-math & 106 &3,602& 85K\\
    \Closure & Google Closure compiler & 133 &7,927&90K\\
    \hline
    Total &  & 357 & 20,109 & 321K\\
    \hline
  \end{tabular}
  \caption{Defects4J V1.0.0 statistics}\label{tab:d4j}
\end{table} 
The benchmark suites are important for evaluating \apr techniques. We choose the Defects4J (V1.0.0) benchmark suite \cite{bib:just2014defects4j}, since it contains hundreds of real-world bugs from real-world systems, and has become the most widely studied dataset for program repair or even software debugging in general~\cite{ghanbari2019practical, jiang2018Shaping, wen2018context, li2019deepfl}.
Table~\ref{tab:d4j} presents the statistics for the Defects4J dataset. Column ``Sub.'' presents the project IDs within Defects4J, while Column ``Name'' presents the actual project names. Column ``\#Bugs'' presents the number of bugs collected from real-world software development for each project, while Columns ``\#Tests'' and ``LoC'' present the number of tests (i.e., JUnit test methods) and the lines of code for the {\tt HEAD} buggy version of each project. 

\subsection{Studied Repair Tools}
Being a well-developed field, APR offers us a cornucopia of choices to select from. According to a recent study~\cite{liu2020efficiency}, there are 31 APR tools targeting Java programs considering two popular sources of information to identify Java APR
tools: the community-led \texttt{\url{program-repair.org}} website and the living
review of APR by Monperrus \cite{bib:monperrus2018living}. 17 of those Java \apr tools are found to be publicly available and applicable to the widely used Defects4J benchmark suite (without additional manually collected information, e.g., potential bug locations) as of July 2019. Note that all such tools are source-level \apr, since the only bytecode-level \apr tool \prapr was only available after July 2019. Table~\ref{tab:tools} presents all such existing Java-based
APR tools, which can be categorized into three main categories according to prior work~\cite{liu2020efficiency}: heuristic-based~\cite{goues2012genprog,jiang2018Shaping,bib:liu2018mining},
constraint-based~\cite{xuan2017nopol,durieux2016dynamoth}, and template-based~\cite{bib:liu2019tbar,wen2018context} repair techniques. In this work, we aims to speed up all existing source-level \apr techniques via on-the-fly patch validation. Therefore, we select one representative \apr tool from each of the three categories for our evaluation to demonstrate the general applicability of our \uniapr framework. All the three considered \apr tools, i.e., \acs~\cite{xiong2017precise}, \simfix~\cite{jiang2018Shaping}, and \capgen~\cite{wen2018context} are highlighted in bold font in the table. 
For each of the selected tools, we evaluate them on all the bugs that have been reported as fixed (with correct patches) by their original papers to evaluate: (1) \uniapr effectiveness, i.e., how much speedup \uniapr can achieve on those tools, and (2) \uniapr precision, i.e., whether the previously fixed bugs are still fixed when running with \uniapr.
\Comment{Please note that we not all of the bugs listed in the table are used in our experiments. Specifically, we have applied the tools only on the bugs that the original papers have reported correct fixes for them. }

\begin{table}[]
\scriptsize
\small
    \centering
    \resizebox{\columnwidth}{!}{\begin{tabular}{|r|p{7cm}|}
        \cline{1-2}
         Tool Category & Tools \\
         \cline{1-2}\hline\hline
         Constraint-based & \textbf{\acs{}}, \nopol, \cardumen{}\Comment{please double check this}, \dynamoth{} \\
         \hline
         Heuristic-based &\textbf{\simfix{}}, \arja, \genprog{}, \jgenprog{}, \jkali{}, \jmutrepair{}, \kali{}, \rsrepair{}  \\
         \hline
         Template-based & \textbf{\capgen}, \tbar{}, \avatar{}, \fixminer{}, \kpar{} \\
         \cline{1-2}
    \end{tabular}}
    \caption{Available Java APR tools for Defects4J}
    \label{tab:tools}
\end{table}

\subsection{Implementation}
\Comment{Implementing a robust system like \uniapr is a difficult engineering undertaking.}

\begin{figure}
\begin{lstlisting}[language=XML, basicstyle=\ttfamily\scriptsize]
      <plugin>
        <groupId>anonymized</groupId>
        <artifactId>uniapr-plugin</artifactId>
        <version>1.0-SNAPSHOT</version>
      </plugin>
\end{lstlisting}
\caption{\label{fig:plugin} \uniapr POM configuration}
\end{figure}

\uniapr has been implemented as a\omitlink{ publicly available} fully automated Maven Plugin\omitlink{~\cite{uniapr}}, on which one can easily integrate any patch generation add-ons. The current implementation involves over 10K lines of Java code.
\Comment{We have implemented three different flavors of \uniapr through more than 10K lines of Java code. The system is available in the form of a Maven plugin on which one can install a patch generation strategy add-on. We have constructed add-ons corresponding to three different APR techniques: \capgen, \simfix, and \acs. \uniapr provides the users with appropriate interface through which the can construct add-ons for other APR tools [?].}
As a Maven plugin, the users simply need to add the necessary plugin information into the POM file. The plugin information can be as simple as shown in Figure~\ref{fig:plugin}. In this way, once the users fire command
\hytt{mvn [anonymized-groupId]:prf-plugin:validate}, the plugin will automatically obtain all the necessary information for patch validation. It will automatically obtain the test code, source code, and 3-rd party libraries from the underlying POM file for the actual test execution. Furthermore, it will automatically load all the patches from the default \texttt{patches-pool} directory (note that the patch directory name and patch can be configured through POM as well) created by the \apr add-ons for patch validation.  The \apr add-ons are constructed by modifying the behavior of the studied APR tools (either through direct modification of the source code or via inheritance and/or decoration when the source code is not available) to not to perform patch validation after generating and/or compiling the patches. \uniapr assumes the patch directory generated by the \apr add-ons to include all available patches represented by their patched bytecode files. Note that, each patch may involve more than one patched bytecode file, e.g., some \apr tools (such as \simfix~\cite{jiang2018Shaping}) can fix bugs with multiple edits.

During patch validation, our system marks the status of all the patches to \texttt{UNKNOWN}. It forks a JVM and passes all the information about the test suites and the subject programs to the child process. The process runs tests on each patch and reporting their status. Note that following the common practice in \apr and the original setting of the studied \apr tools~\cite{jiang2018Shaping, xiong2017precise, wen2018context}, the process always runs the originally failing tests earlier than the originally passing tests for each patch. The reason is that the originally failing tests have a high probability to fail again and can falsify non-plausible patches faster.  The execution results can be either of the following: \texttt{PLAUSIBLE} if the patch passes all the tests, \texttt{NON-PLAUSIBLE} if the the patch fails to pass a test, \texttt{TIME\_OUT} if the patch times out on some test, \texttt{MEMORY\_ERROR} if the patch runs out of heap space, and \texttt{UNKNOWN\_ERROR} if testing the patch makes the child JVM to crash. We use TCP Socket Connections as a means of communication between processes. \uniapr repeats this process of forking and receiving report results until all the patches are executed. It is worth noting that it is very well possible to fork two or more processes to take maximum advantage of today's powerful computers' potentials. However, for a fair comparison with existing work, we always ensure that only one JVM is running patch validation at any given time stamp.\Comment{Please note further that in the experiments described in the rest of this paper, we have used a variant of \uniapr that does not apply any isolation in the forked JVM for patch validation.}

\subsection{Experimental Setup}

For each of the studied \apr tools, we perform the following experiments on all the bugs that have been reported as fixed in their original papers. First, we execute the original \apr tools to trace their original patch validation time and detailed repair results (e.g., the number of patches executed and plausible patches produced). Next, we modify the studied tools and make them conform to \uniapr add-on interfaces, i.e., dumping all the generated patches into the patch directory format required by \uniapr. Then, we launch our \uniapr to validate all the patches generated by each of the studied \apr tools, and trace the new patch validation time and detailed repair results. \Comment{Note that for each of the studied tools, we run their original patch validation setting. }

To evaluate our \uniapr system, we include the following metrics: (1) the speedup compared with the original patch validation time, measuring the effectiveness of \uniapr, and (2) the repair results compared with the original patch validation, measuring the precision of our patch validation (i.e., checking whether \uniapr fails to fix any bugs that can be fixed via traditional patch validation).

\Comment{For each of the studied APR tool, along with the add-on that generates patches by applying the algorithm implemented in the APR tool and compiling them, we have implemented a variant of each tool that prints out the time elapsed during running patch validation using vanilla testing. Some tools does vanilla testing by invoking a build system while directly call JUnit runner from command-line. In either case, we have guarded the code for patch validation with a stopwatch code so as to calculate the time elapsed during patch validation. We then use Unix's \texttt{grep} tool to gather and tally the elapsed times to calculate overall time spent on patch validation via vanilla testing. We then apply \uniapr on the same buggy program to calculate the time elapsed between patch validation starts until all patches have some status not equal to \texttt{UNKNOWN}. We use these two numbers to calculate the speedup gain.

Lastly, we compare the number of plausible patches reported by the original tools and \uniapr to report the differences, if any. Please note that we use patch ids described in the previous section to make sure that plausible patches reported by \uniapr are compatible with that of original tools.\lingming{we should also add that when the plausible patch numbers are the same, we further manually ensured that they are the same plausible patch}
}
All our experimentation is done on a Dell workstation with Intel Xeon CPU E5-2697 v4@2.30GHz and 98GB RAM, running Ubuntu 16.04.4 LTS and Oracle Java 64-Bit Server version 1.7.0\_80.

\section{Result Analysis}
\label{sec:res}
In this section we present the detailed result analysis for the research questions outlined in Section \ref{sec:rqs}.

\subsection{RQ1: Results for Vanilla On-the-fly Patch Validation}

\begin{figure*}
        \begin{subfigure}[b]{0.33\textwidth}
                \includegraphics[width=\linewidth]{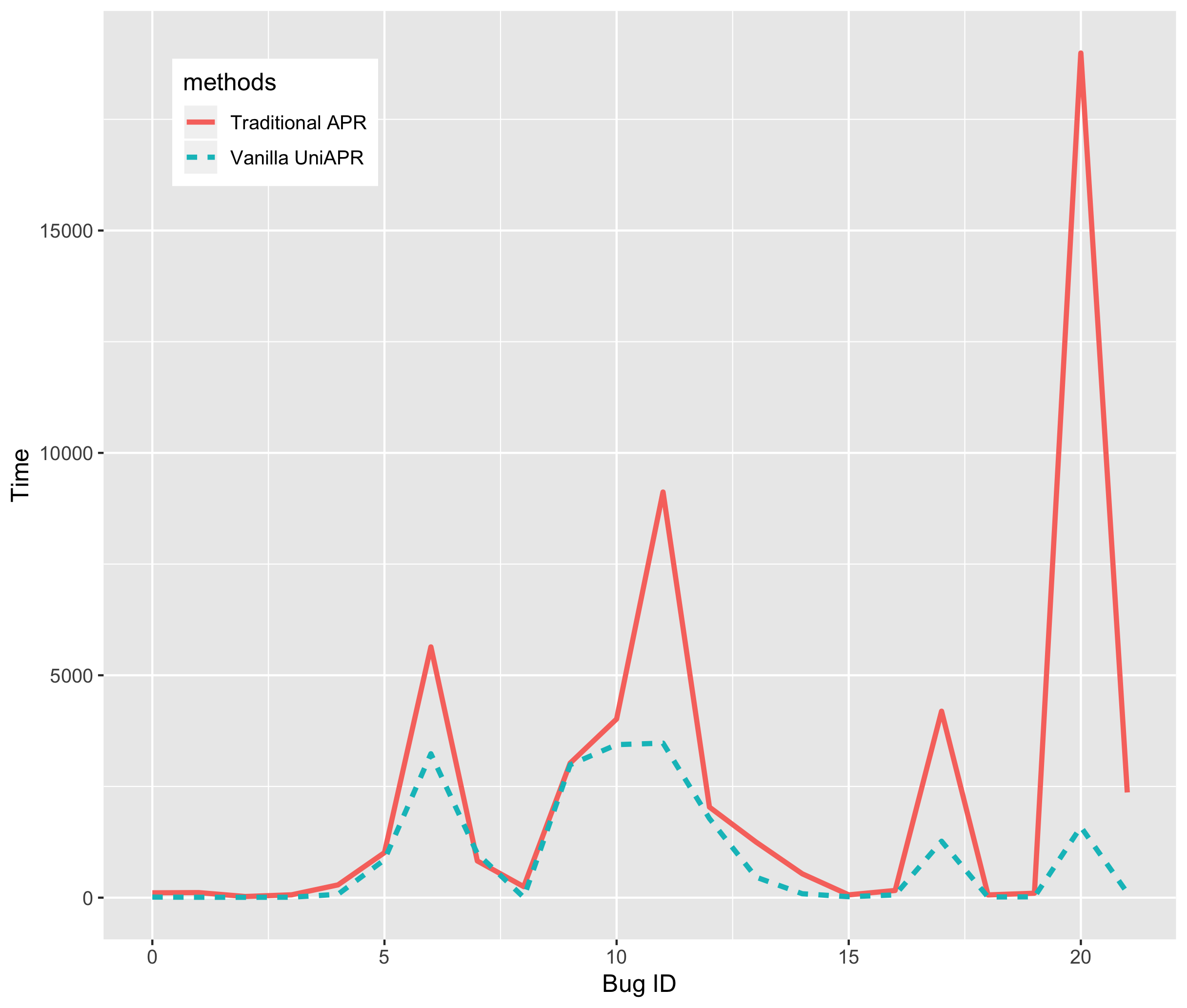}
                \caption{CapGen}
                \label{fig:capgen_original_pit}
        \end{subfigure}%
        \begin{subfigure}[b]{0.33\textwidth}
                \includegraphics[width=\linewidth]{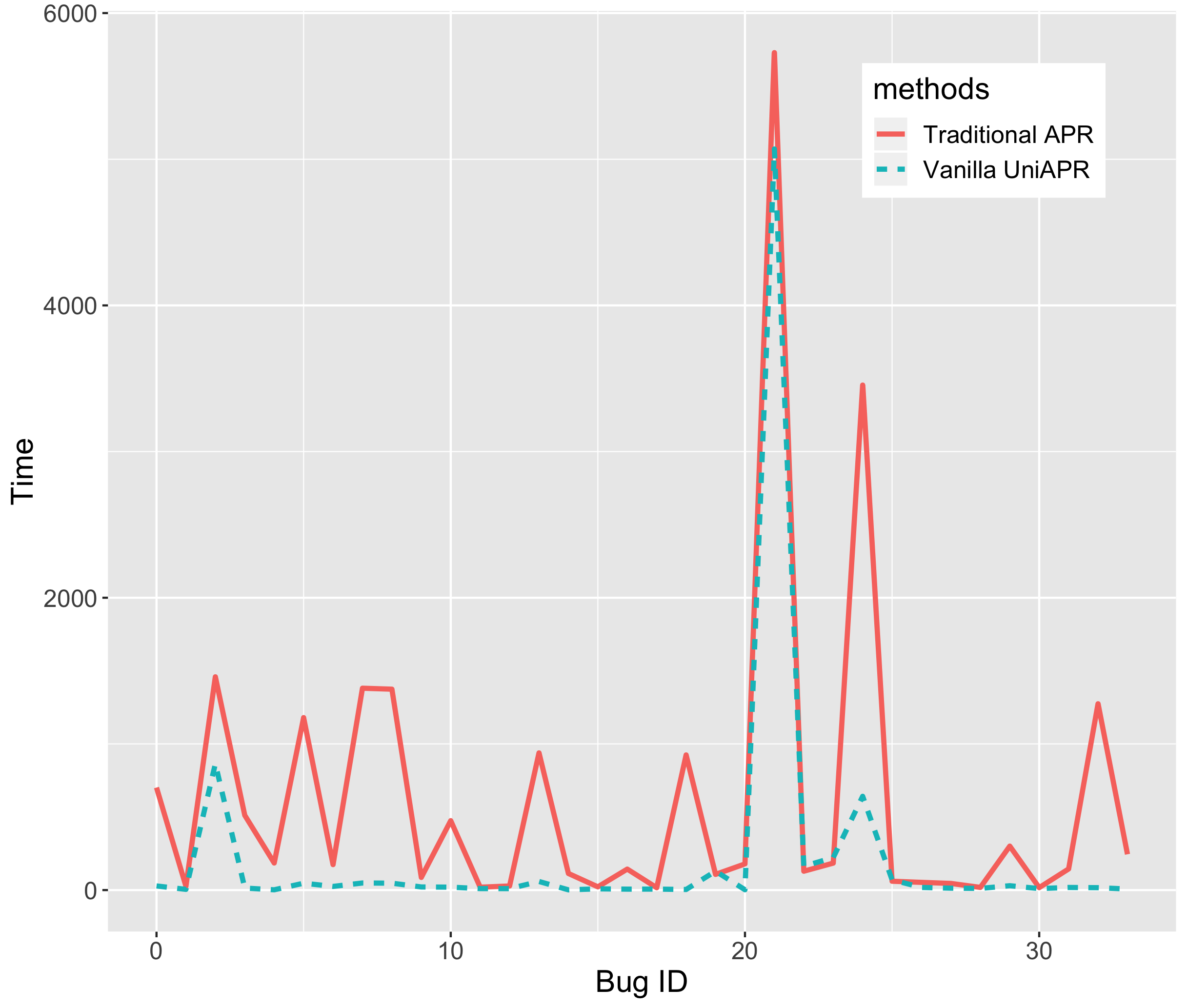}
                \caption{SimFix}
                \label{fig:simfix_original_pit}
        \end{subfigure}%
        \begin{subfigure}[b]{0.33\textwidth}
                \includegraphics[width=\linewidth]{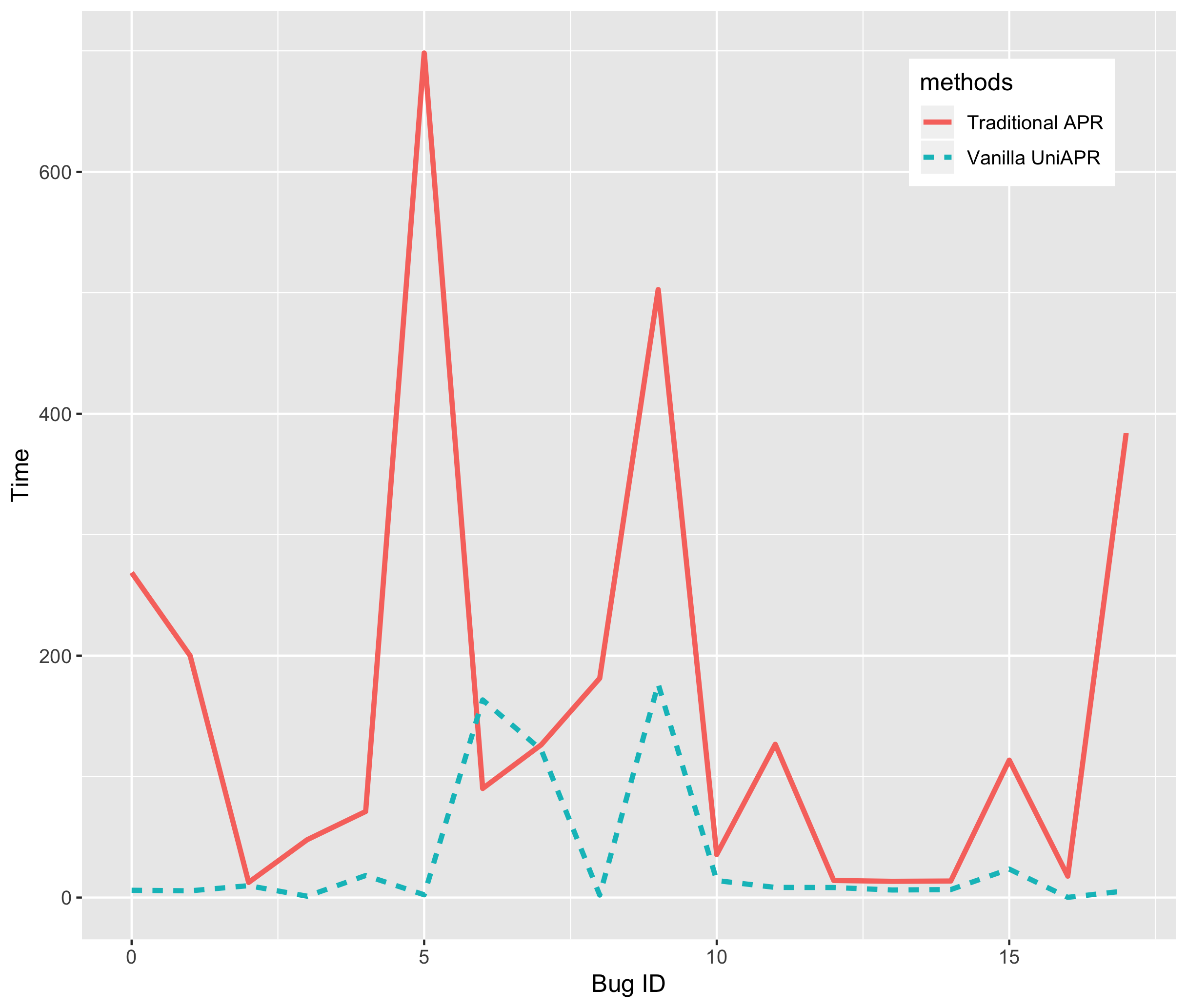}
                \caption{ACS}
                \label{fig:acs_original_pit}
        \end{subfigure}%
        
        \caption{Speedup achieved by vanilla \uniapr}
        \label{fig:speedup}
\end{figure*}

\Comment{In order to answer RQ1, we executed latest versions of APR tools \capgen (representing mutation-/pattern-based techniques), \simfix (representing grafting-based techniques), and \acs (representing semantic-based techniques) on the bugs that they can generate correct fixes. We recorded patch generation time as well as the number of plausible patches that each tool generated.}
\subsubsection{Effectiveness}
For answering this RQ, we executed vanilla \uniapr (without JVM-reset) that is configured to use the add-on corresponding to each studied APR tool. Please note that we measure the original patch validation time by each original \apr tool, and compare that against the patch validation time using vanilla \uniapr. The only exception is for \capgen: we observed that \capgen performs much faster even than vanilla \uniapr for some cases; digging into the 
decompiled \capgen code (the \capgen source code is not available), we realized that the \capgen tool excluded some (expensive) tests for certain bugs (confirmed by the authors). Therefore, to enable a fair comparison, for \capgen, we build a variant for our \uniapr framework that simply restarts a new JVM for each patch. The main experimental results are presented in Figure~\ref{fig:speedup}. In each sub-figure, the horizontal axis presents all the bugs that have been reported to be fixed by each studied tool, while the vertical axis presents the time cost (s); the solid and dashed lines then present the time cost for traditional patch validation and our vanilla \uniapr, respectively.
From the figure, we can have the following observations:

First, for all the studied \apr tools, vanilla \uniapr can substantially speed up the existing patch validation component for all state-of-the-art \apr tools. For example, when running \capgen on Math-80, the traditional patch validation costs 18,991s while on-the-fly patch validation via vanilla \uniapr takes only 1,590s to produce the same patch validation results, i.e., 11.9X speedup; when running \simfix{} on Closure-62, the traditional patch validation costs 1,381s, while on-the-fly patch validation via vanilla \uniapr takes only 48s to produce the same patch validation results, i.e., 28.8X speedup; when running \acs{} on Math-82, the traditional patch validation costs 127s, while on-the-fly patch validation via vanilla \uniapr takes less than 9s to produce the same patch validation results, i.e., 14.3X speedup. To the best of our knowledge, this is the first study demonstrating that on-the-fly patch validation can also substantially speed up the existing state-of-the-art source-level \apr techniques. 

\begin{figure}
    \centering
    \includegraphics[width=6.5cm, height=4cm]{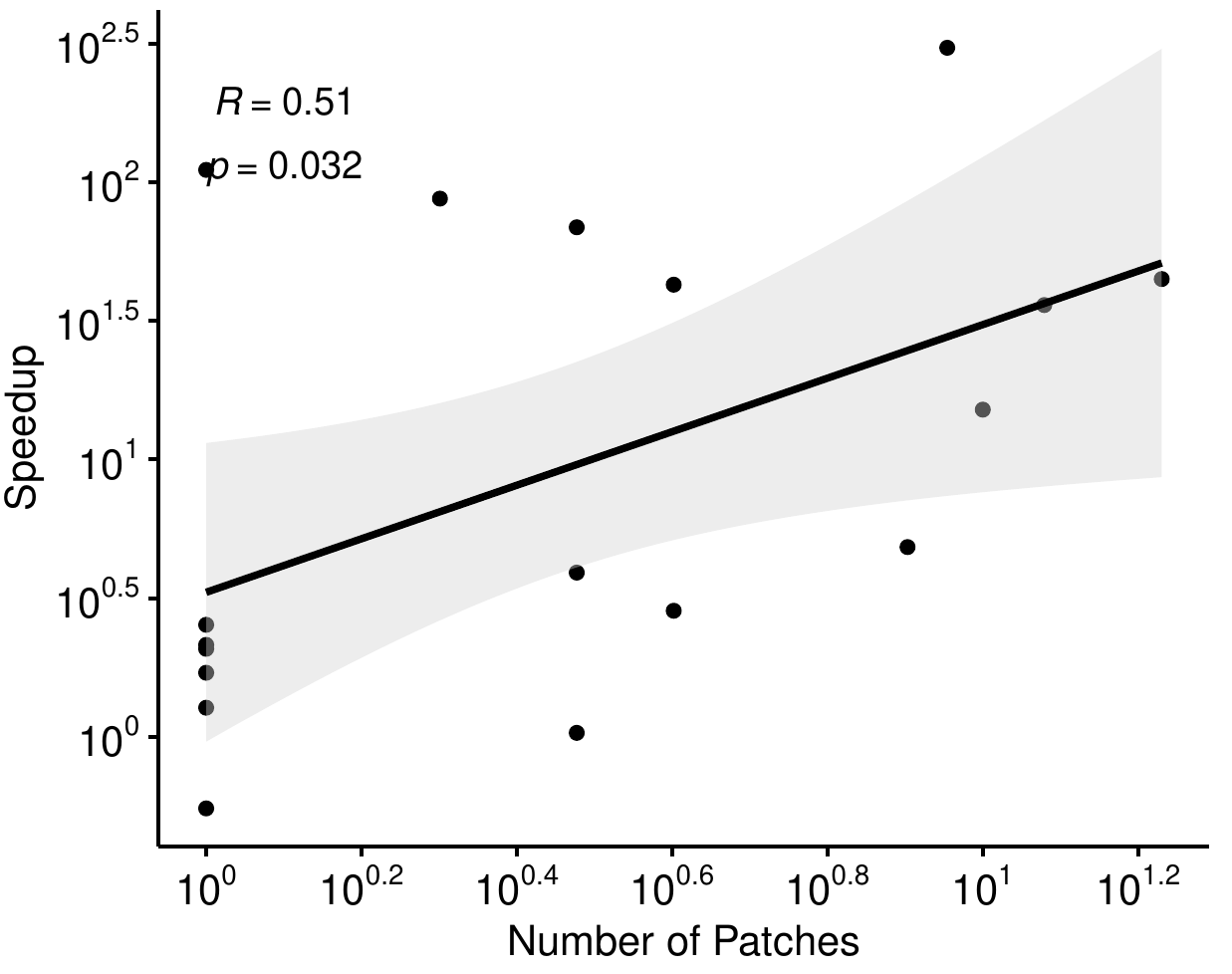}
    \caption{Correlation between patch number and speedup}
    \label{fig:corr}
\end{figure}
Second, while we observe clear speedups for the vast majority of the bugs, the achieved speedups vary a lot for all the studied \apr tools on all the studied bugs. The reason is that the speedups are impacted by many different factors, such as the number of patches executed, the number of bytecode files loaded for each patch execution, the individual test execution time, and so on. For example, we observe that \uniapr even slows down the patch validation for \acs slightly on one bug (i.e., for 1min). Looking into the specific bug (i.e., Math-3), we find that \acs only produces one patch for that bug, and there is no JVM sharing optimization opportunity for \uniapr on-the-fly patch validation. To further confirm our finding, we perform the Pearson Correlation Coefficient analysis~\cite{pearson1895notes} (at the significance level of 0.05) between the number of patches for each studied bug and its corresponding speedup for \acs. Shown in Figure~\ref{fig:corr}, the horizontal axis denotes the number of patches, while the vertical axis denotes the speedup achieved; each data point represents one studied bug for \acs. From this figure, we can observe that \uniapr tends to achieve significantly larger speedups for bugs with more patches (at the significance level of 0.05), demonstrating that \uniapr tends to achieve larger speedups for larger systems with more patches.

\begin{tcolorbox}
\textbf{Finding-1:} This study demonstrates for the first time that vanilla on-the-fly patch validation can also substantially speed up the existing state-of-the-art source-level \apr techniques; furthermore, \uniapr tends to perform better for systems/bugs with more patches, demonstrating the scalability of on-the-fly patch validation.
\end{tcolorbox}

\subsubsection{Precision.}
\label{sec:prec}
\begin{table}
    \centering
    \small
    \begin{tabular}{l|c|c|r}
    \hline
         Tool& \# All&\# Mismatch&Ratio (\%)  \\\hline
         \capgen&22&3&13.64\%\\         
         \simfix&34&1&2.94\%\\         
         \acs&18&0&0.00\%\\   \hline      
         All&74&4&\unsafeRatio\\
         \hline
    \end{tabular}
    \caption{Inconsistent fixing results}
    \label{tab:unfixed}
\end{table}
We further study the number of bugs that vanilla \uniapr does not produce the same repair results as the traditional patch validation (that restarts a new JVM for each patch). Table~\ref{tab:unfixed} presents the summarized results for all the studied \apr tools on all their fixable bugs. In this table, Column ``Tool'' presents the studied \apr tools, Column ``\# All'' presents the number of all studied fixable bugs for each \apr tool, Column ``\# Mismatch'' presents the number of bugs that vanilla \uniapr has inconsistent fixing results with the original \apr tool, and Column ``Ratio (\%)'' presents the ratio of bugs with inconsistent results. From this table, we can observe that vanilla \uniapr produces imprecise results for \unsafeRatio of the studied cases overall. To our knowledge, this is the first empirical study demonstrating that on-the-fly patch validation may produce imprecise/unsound results compared to traditional patch validation. Another interesting finding is that 3 out of the 4 cases with inconsistent patching results occur on the \capgen \apr tool. One potential reason is that \capgen is a pattern-based \apr system and may generate far more patches than \simfix and \acs. For example, \capgen on average generates over 1,400 patches for each studied bug, while \simfix only generates around 150 on average. In this way, \capgen has way more patches that may affect the correct patch execution than the other studied \apr tools. Note that \simfix has only around 150 patches on average since we only studied its fixed bugs, if we had considered the unfixed bugs as well, \simfix will produce many more patches, exposing more imprecise/unsound patch validation issues as well potentially leading to larger \uniapr speedups.

\begin{tcolorbox}
\textbf{Finding-2:} This study presents the first empirical evidence that vanilla on-the-fly patch validation does incur the imprecise/unsound patch validation issue, e.g., failing to fix \unsafeRatio of the studied cases.
\end{tcolorbox}

\Comment{The add-on stores each patched source file before compilation and skips validation. It then invoked \texttt{javac} to compile each patch and dumps the resulting class file(s) into a directory corresponding to each patch. In this way, it constructs \texttt{patches-pool} directory for each bug. Please note that we directly compile the patches so that we can have more control on the class files generated as the result of compilation. We opted for this method as a class might have several subclasses and the patching might occur inside a subclass and/or an anonymous class. After obtaining \texttt{patches-pool} directory for all the bugs, we record the time \uniapr spends validating the patches via HotSwap technique.}

\Comment{Tables \ref{tab:capgenRes}, \ref{tab:simfixRes}, and \ref{tab:acsRes} summarize the results for comparing \uniapr with \capgen, \simfix, and \acs, respectively. In the tables, the column ``BugId'' represents the identifier for a Defects4J bug, e.g. \Math-33 denotes bug \#33 from \Math project in Defects4J. The column ``Ori.(s)'' represents the time (in seconds) each APR tool spends on validating the patches through their specific way of vanilla testing (which could be invoking a build system, Defects4J's interface, or directly using JUnit's command-line interface), while Column ``\uniapr(s)'' represents that of \uniapr. ``Gain'' represents the amount of speed-up that \uniapr offers in case of each bug. Column ``Def.'' represents the number of plausible patches that \uniapr finds via HotSwap trick, while the column ``Vanil.'' represents the number of plausible patches that \uniapr finds after it falls back to degenerate method of vanilla testing. Lastly, the columns ``Pub.'' and ``Impl.'' represent the number of expected plausible parches reported in the original paper and those that we obtained by reproducing the experiments, respectively.}

\Comment{We observed that in \percentTimeOutDelay of the studied bugs, \uniapr takes more time than vanilla testing used by the tools. These are the bugs that have many patches generated and a large portion of the patches \emph{hang} until they time out and get killed by the framework. Since we have designed \uniapr as a general patch validation framework that is intended to work across different varieties of bugs and patch generation strategies, we cannot optimize the time out value for a specific benchmark. Instead, we have to follow a generic approach like the one outlined in Section \ref{sec:discussion}.}

\subsection{RQ2: Results for On-the-fly Patch Validation via JVM-Reset}

\begin{figure}
    \centering
    \includegraphics[width=0.35\textwidth, height=0.24\textwidth]{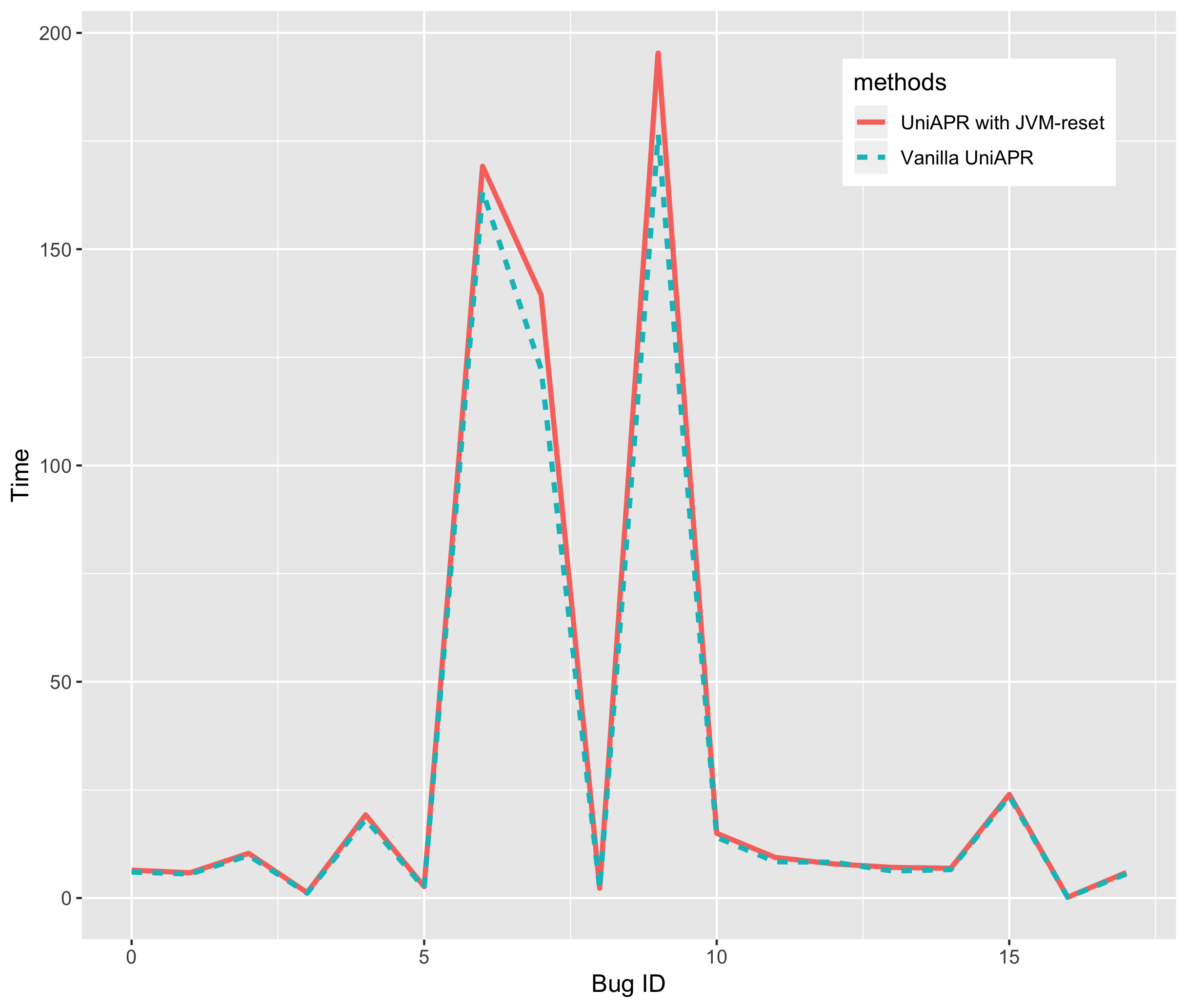}
    \caption{JVM-reset overhead on UniAPR}
    \label{fig:reset-time}
\end{figure}
\subsubsection{Effectiveness}
We now present the experimental results for our \uniapr{} with JVM-reset. We observe that \uniapr with JVM reset has negligible overhead compared with the vanilla \uniapr on all the studied bugs for all the studied \apr systems. Figure~\ref{fig:reset-time} presents the time cost comparison among the two \uniapr variants on \acs (note that we omit the results for the other two \apr tools since they have even lower average overhead). In the figure, the horizontal axis presents all the bugs studied by \acs while the vertical axis presents the time cost; the solid and dashed lines present the time cost for \uniapr with JVM-reset and our vanilla \uniapr. Shown in the figure, JVM reset has incurred negligible overhead among all the studied bugs for \acs on \uniapr, e.g., on average 
8.33\% overhead. The reason is that class reinitializations only need to be performed at certain sites for only the classes with pollution sites. 
Also, we have various optimizations to speed up JVM reset. For example, although our basic JVM-reset approach in Figure~\ref{fig:reset} performs runtime checks on a concurrent \codeIn{HashMap}, our actual implementation uses arrays for faster class status tracking/check. Furthermore, we observe that the overhead does not change much regardless of the bugs studied, indicating that our \uniapr with JVM-reset has stable overhead across different systems/bugs.

\begin{tcolorbox}
\textbf{Finding-3:} \uniapr with JVM-reset only incurs negligible overhead (e.g., less than \resetOver{} for all studied tools) compared to the vanilla \uniapr, demonstrating the scalability of \uniapr with JVM-reset.
\end{tcolorbox}

\subsubsection{Precision} 
According to our experimental results, \uniapr with JVM-rest is able to produce exactly the same \apr results as the traditional patch validation, i.e., \uniapr with JVM-reset successfully fixed all the bugs that vanilla \uniapr failed to fix. We now discuss all the four bugs that \uniapr{} can fix while vanilla \uniapr{} without JVM reset cannot fix in details: 

\begin{figure}[h!]
\center
\begin{lstlisting}[language=JAVA, basicstyle=\ttfamily\scriptsize]
// org.apache.commons.lang3.StringEscapeUtilsTest.java
    public void testUnescapeHtml4() {
        for (int i = 0; i < HTML_ESCAPES.length; ++i) {
            String message = HTML_ESCAPES[i][0];
            String expected = HTML_ESCAPES[i][2];
            String original = HTML_ESCAPES[i][1];
            // assertion failure: ampersand expected:<bread &[] butter> but was:<bread &[amp;] butter>
            assertEquals(message, expected, StringEscapeUtils.unescapeHtml4(original)); 
    ...
\end{lstlisting}
\caption{\label{fig:lang6} Test failed for a plausible patch without JVM-reset on Lang-6}
\end{figure}

Figure~\ref{fig:lang6} presents the test that fails on the only plausible (also correct) patch of Lang-6 (using \capgen) when running \uniapr{} without JVM-reset. Given the expected resulting string \codeIn{``bread \&[] butter''}, the actual returned one is \codeIn{``bread \&[amp;] butter''}.  Digging into the code, we realize that class \codeIn{StringEscapeUtils} has a static field named \codeIn{UNESCAPE\_HTML4}, which is responsible for performing the \codeIn{unescapeHtml4} invocation. However, during earlier patch executions, the actual object state of 
that field is changed, making the \codeIn{unescapeHtml4} method invocation return problematic result with vanilla \uniapr{}. In contrast, when running \uniapr{} with JVM-reset, field \codeIn{UNESCAPE\_HTML4} will be recreated before each patch execution (if accessed) and will have a clean object state for performing the \codeIn{unescapeHtml4} method invocation. 

\begin{figure}[h!]
\center
\begin{lstlisting}[language=JAVA, basicstyle=\ttfamily\scriptsize]
// org.apache.commons.math3.EventStateTest.java
    public void testIssue695() {
        FirstOrderDifferentialEquations equation = new FirstOrderDifferentialEquations() {
        ...
        double tEnd = integrator.integrate(equation, 0.0, y, target, y);
        ...

    private static class ResettingEvent implements EventHandler {
        private static double lastTriggerTime = Double.NEGATIVE_INFINITY;
        public double g(double t, double[] y) {
            // assertion error
            Assert.assertTrue(t >= lastTriggerTime);
            return t - tEvent;
        }
        ...
\end{lstlisting}
\caption{\label{fig:math30&math41} Test failed for a plausible patch without JVM-reset on Math-30 and Math-41}
\end{figure}
Figure~\ref{fig:math30&math41} shows another test that fails on the only plausible (and correct) patch of Math-30 when running vanilla \uniapr (without JVM-reset) with \capgen patches as well as the only plausible (and correct) patch of Math-41 when running vanilla \uniapr with \simfix patches. Looking into the code, we find that the invocation of \codeIn{integrate()} in the test will finally call the method \codeIn{g()} in class \codeIn{ResettingEvent} (in the bottom). The static field \codeIn{lastTriggerTime} of class \codeIn{ResettingEvent} should be \codeIn{Double.NEGATIVE\_INFINITY} in Java, which means the assertion should not fail. Unfortunately, the earlier patch executions pollute the state and change the value of the field. Thus, the test failed when running with vanilla \uniapr{} on the two plausible patches. In contrast, \uniapr with JVM-reset is able to successfully recover the field value.

\begin{figure}
\center
\begin{lstlisting}[language=JAVA, basicstyle=\ttfamily\scriptsize]
// org.apache.commons.math3.genetics.UniformCrossoverTest.java
    public class UniformCrossoverTest {
        private static final int LEN = 10000;
        private static final List<Integer> p1 = new ArrayList<Integer>(LEN);
        private static final List<Integer> p2 = new ArrayList<Integer>(LEN);
        public void testCrossover() {
            performCrossover(0.5);
            ...
            
        private void performCrossover(double ratio) {
            ...
            // assertion failure: expected:<0.5> but was:<5.5095>
            Assert.assertEquals(1.0 - ratio, Double.valueOf((double) from1 / LEN), 0.1);
            ...
\end{lstlisting}
\caption{\label{fig:t1_math5} Test1 failed for a plausible patch without JVM-reset on Math-5}
\end{figure}

\begin{figure}
\center
\begin{lstlisting}[language=JAVA, basicstyle=\ttfamily\scriptsize]
// org.apache.commons.math3.complex.ComplexTest.java
    public class ComplexTest {
        private double inf = Double.POSITIVE_INFINITY;
        ...
        public void testMultiplyNaNInf() {
            Complex z = new Complex(1,1);
            Complex w = z.multiply(infOne);
            // assertion failure: expected:<-Infinity> but was:<Infinity>
            Assert.assertEquals(w.getReal(), inf, 0);
            ...
\end{lstlisting}
\caption{\label{fig:t2_math5} Test2 failed for a plausible patch without JVM-reset on Math-5}
\end{figure}

There are four plausible \capgen patches on Math-5 (one is correct) when running with the traditional patch validation. With vanilla \uniapr, all the plausible patches failed on some tests. Figure~\ref{fig:t1_math5} shows the test that fails on three plausible patches (including the correct one) on Math-5. The expected value of the assertion should be 0.5, but the actual value turned to 5.5095 due to the change of variable \codeIn{from1}. After inspecting the code, we found the value of \codeIn{from1} is decided by two static fields \codeIn{p1} and \codeIn{p2} in class \codeIn{UniformCrossoverTest}. The other earlier patch executions pollute the field values, leading to this test failure when running with vanilla \uniapr. Figure~\ref{fig:t2_math5} presents another test that fails on one plausible patch on Math-5. The expected value from invocation \codeIn{w.getReal()} should be \codeIn{Infinity}, which should be the same as field \codeIn{inf} defined in class \codeIn{ComplexTest}; however, the actual result from the method invocation is \codeIn{-Infinity}. The root cause of this test failure is similar to the previous ones, the static fields \codeIn{NaN} and \codeIn{INF} in class \codeIn{Complex} are responsible for the result of method invocation \codeIn{getReal()}. In this way, \codeIn{getReal()} returns a problematic result because the earlier patch executions changed the corresponding field values. In contrast, using \uniapr with JVM-reset, all the four plausible patches are successfully produced.

\begin{tcolorbox}
\textbf{Finding-4:} \uniapr with JVM-reset is able to successfully fix all the studied bugs, i.e., mitigating the imprecise/unsound patch validation issue by vanilla on-the-fly patch validation.
\end{tcolorbox}
\subsection{Discussion}\label{sec:discussion}
\Comment{\lingming{the timeout discussion in this section is not helping the main goal of this paper; instead, it makes reviewers begin to think about whether the speedup is due to on-the-fly patching or due to the specific timeout policy (which could also be easily done by existing tools); I would rather discuss some unresolved issues (even using JVM reset) of on-the-fly patching.}\ali{I cannot see why the issue of timeout, if discussed properly, could be a problem.}\lingming{it's fine, we can come back later, you can go ahead with other sections (such as related work ) for now.}}

Having single JVM session for validating more than one patch has the immediate benefit of skipping the costly process creation, validations done by the JVM, and Just-in-Time compilation. As per our experiments, this offers up to several hundred times speed up in patch validation. On the other hand, this approach might have the following limitations: 

First, the execution of the patches might interfere with each other, i.e. the execution of some tests in one patch might have side-effects affecting the execution of other tests on another patch. \uniapr mitigates these side-effects by resetting static fields to their default values and resetting JDK properties. Although our experimental results demonstrate that such JVM reset is able to fix all bugs fixed by the traditional patch validation, resetting in-memory state of JVM might not be enough as the side-effects could propagate via operating system or the network.\Comment{We will discuss the ramifications of such side-effects shortly.} Our current implementation provides a public interface for the users to resolve such issue between patch executions (note that no subject systems in our evaluation require such manual configuration). In the near future, we will study more subject programs to fully investigate the impact of such side effects and design new solutions to address them fully automatically.

Second, HotSwap-based patch validation does not support patches that involve changing the layout of the class, e.g. adding/removing fields and/or methods to/from a class. It also does not support patches that occur inside non-static inner classes, anonymous classes, and lambda abstraction. Luckily, the existing \apr techniques mainly target patches within ordinary method bodies, and our \uniapr framework is able to reproduce all correct patches for all the three studied state-of-the-art techniques. Another thing that worths discussion is that HotSwap originally does not support changes in static initializers; interestingly, our JVM-reset approach naturally helped \uniapr to overcome this limitation, since the new initializers can now be reinvoked based on our bytecode transformation to reinitialize the classes. In the near future, we will further look into other more advanced dynamic class redefinition techniques for implementing our on-the-fly patch validation, such as JRebel~\cite{jrebel} and DCEVM~\cite{dcevm}.

\section{Related Work}
\label{sec:related}
In this section, we first talk about the related work in the automated program repair (\apr) area, including techniques for reducing \apr cost (Section~\ref{sec:aprRelated}). Next, we further talk about the cost reduction techniques in the mutation testing area (Section~\ref{sec:mut}), which shares similar cost issues with \apr.

\subsection{Automated Program Repair}
\label{sec:aprRelated}
APR has been the subject of intense research in the last decade \cite{goues2019automated,gazzola2017automatic,monperrus2018automatic}. APR techniques can be classified into two broad families: (1) techniques that monitor the execution of a system to find deviations from certain specifications, and then heal the system by modifying its runtime state in case of any abnormal behaviors \cite{perkins2009automatically,long2014automatic}, and (2) so-called generate-and-validate (\gv) techniques that attempt to fix the bug by first generating a pool of patches (at the code representation level) and validating the patches via certain rules and/or checks \cite{goues2012genprog,nguyen2013semfix,wen2018context,ghanbari2019practical,jiang2018Shaping,martinez2016ASTOR, lou2019history, jiang2019inferring}.
\gv techniques have been widely studied in recent years, since it can substantially reduce developer efforts in both automated~\cite{goues2012genprog} and manual~\cite{lou2020can} bug fixing for improving software productivity. To date, researchers have designed various \gv \apr techniques based on heuristics \cite{bib:liu2018mining,goues2012genprog,jiang2018Shaping}, constraint solving~\cite{xuan2017nopol,durieux2016dynamoth,nguyen2013semfix,mechtaev2016angelix},\Comment{\ali{ I believe there is a better classification, e.g. pre-defined templates are in fact heuristic as they are either mined from real-world commits or are based on mutations developed over years based on common sense and experience }\lingming{it is ok: we just want to justify our selection of three tools, and we followed prior work on apr categorization ~\cite{liu2020efficiency}}} and pre-defined templates~\cite{kim2013automatic,ghanbari2019practical,bib:liu2019tbar}.
\emph{Heuristic-based \apr techniques} investigate various strategies to explore a (potentially infinite) search space of syntactic program changes. GenProg~\cite{goues2012genprog}, the pioneering work for \apr, leverages genetic programming to compose and mutate single-change patches into more advanced ones that can fix more complex bugs. Later on, RSRepair~\cite{qi2014strength} demonstrates that using random search rather than genetic programming can help GenProg to mitigate the search space explosion problem. Recently, \simfix~\cite{jiang2018Shaping} 
and ssFix \cite{xin2017leveraging} leverage code search information (e.g., from the current project under test or even other projects) to help further reduce the potential search space. 
\emph{Constraint-based \apr techniques} generally leverage advanced constraint-solving or synthesis techniques to fix certain types of bugs. For example, ACS~\cite{xiong2017precise}, Nopol~\cite{xuan2017nopol}, and Cardumen~\cite{martinez2018ultra} aim to fix problematic conditional expressions. Nopol~\cite{xuan2017nopol} transforms the \apr problem into a satisfiability problem and leverages off-the-shelf SMT solvers to find the fixing ingredients.
 Cardumen~\cite{martinez2018ultra} fixes bugs by synthesizing potential correct expression candidates with its mined templates from the
current program under repair.
ACS~\cite{xiong2017precise} leverages various dimensions of information (including local contexts, API document, and code mining information) to directly synthesize the correct conditional expressions.
\emph{Template-based \apr techniques} are also often denoted as pattern-based \apr techniques. The basic idea is to fix program bugs via a set of predefined rules/templates/patterns. The PAR~\cite{kim2013automatic} work produces patches based on a list of fixing patterns manually summarized from a large number of human-written patches. Later on, researchers have also proposed various techniques to automatically mine potential fixing patterns from historical bug fixes, such as the CapGen~\cite{wen2018context} and FixMiner~\cite{koyuncu2020fixminer} work. More recently, the TBar~\cite{bib:liu2019tbar} work presents an empirical study of prior fixing patterns used in the literature.

 Despite this spectacular progress in designing new \apr techniques, very few of the works have attempted to reduce the time cost for \apr, especially the patch validation time which dominates repair process\Comment{, and in fact no research so far has specifically targeted this problem}. For example, \jaid \cite{chen2017contract} uses patch schema to fabricate meta-programs that bundle several patches in a single source file, while \skfix \cite{hua2018towards} uses sketches \cite{lezama2008program} to achieve a similar effect. Although they can potentially help with patch generation and compilation, they still require validating each patch in a separte JVM, and have been shown to be rather costly during patch validation~\cite{ghanbari2019practical}. More recently, \prapr \cite{ghanbari2019practical} uses direct bytecode-level mutation and HotSwap technique to generate and validate patches on-the-fly, thereby bypassing expensive operations such as AST manipulation and compilation on patch generation side and process creation and JVM warm-up on patch validation side. This makes \prapr at least an order of magnitude faster than state-of-the-art \apr (including \jaid and \skfix). However, \prapr is limited to only the bugs that can be fixed via bytecode manipulation, and can also return imprecise patch validation results due to the potential JVM pollution.

Motivated by \prapr work, in this paper, we introduce \uniapr, the first unified \apr framework for all source-code and bytecode level \apr. Compared with \prapr, \uniapr can be applied to any of the existing source-level \apr technique and is not limited by the \prapr bug-fixing patterns at the bytecode level. Furthermore, although \uniapr also uses the JVM HotSwap technique to validate patches without unnecessary restart of JVM sessions, different from \prapr, \uniapr resets JVM internal state between patches so as to contain side effects of patch executions for fast\&precise patch validation. 
Lastly, unlike \prapr, this framework is generic and can use any existing APR technique as a \emph{patch generation add-on}; in this way, it provides a natural platform for combining different APR techniques to take maximum advantage of their strengths.

\subsection{Cost Reduction for Mutation Testing}
\label{sec:mut}
Mutation testing~\cite{ammann2016introduction}, a traditional testing methodology studied for decades, aim to generate artificial bugs to simulate real bugs for test quality evaluation. In recent years, mutation testing has also been applied to various other areas, including regression testing~\cite{lou2015mutation, shi2014balancing}, test generation~\cite{zhang2010test,papadakis2010towards}, bug localization~\cite{li2019deepfl, li2017transforming, zhang2013injecting, papadakis2015metallaxis, moon2014ask}, and even \apr~\cite{ghanbari2019practical, debroy2010Using}. In fact, mutation testing shares similarities with APR in that both involve making small changes to program and repeatedly running tests. The repeated executions of tests on a large number of mutated program versions also dominate the end-to-end time of mutation testing. Therefore, a body of research is devoted to reduce such mutation execution cost~\cite{zhang2018predictive,wangFaster2017,coles2016pit,king1991fortran}.
Weak mutation testing~\cite{howden1982weak}, one of the earliest techniques that attempts to reduce mutation testing cost, executes each mutated program partially (only to the mutated program locations) and checks the program internal state change to reduce the mutation testing time. However, it may produce rather imprecise mutation testing results, since the internal state changes may not always propagate to the end.
Split-stream \cite{king1991fortran} attempts to reduce mutation testing cost by reusing the state before the first mutation point. In a recent work \cite{wangFaster2017}, Wang et al. introduce the concept of Equivalence Modulo States to further elaborate on split-stream technique by reusing the shared states, as much as possible, even after the mutation point. In all these optimization approaches, creating a new process is done as a last resort to avoid the overhead of process creation and to use the shared state as much as possible. However, such techniques cannot handle programs with external resource accesses (e.g., file/network accesses) well~\cite{wangFaster2017}. Process creation overhead in systems like JVM is especially pronounced as virtually all the optimization tasks, as well as Just-in-Time compilation, are done at runtime. Therefore, compared to natively executed programs, JVM-based programs are expected to take relatively longer time to warm up. Therefore, state-of-the-art mutation engine for JVM-based systems, PIT~\cite{coles2016pit}, attempts to use a single process to perform mutation testing instead of creating a separate process for testing each mutation. PIT has been demonstrated to achieve a significant speedup over similar mutation engines (such as MAJOR \cite{just2014major} and Javalanche \cite{schuler2009javalanche}), and represent state of the art.
 In fact, the \prapr \apr tool is built on top the PIT mutation engine. Compared with our \uniapr system, PIT only supports very simple change/mutation patterns at the bytecode level and also cannot handle the JVM state pollution problem.

\section{Conclusion}
\label{sec:conclude}
Automated program repair (\apr) has been extensively studied in the last decade, and various \apr systems/tools have been proposed. However, state-of-the-art \apr tools still suffer from the efficiency problem largely due to the expensive patch validation process. In this work, we have proposed a unified on-the-fly patch validation framework for all JVM-based \apr systems. Compared with the existing on-the-fly patch validation work~\cite{ghanbari2019practical} which only works for bytecode \apr, this work generalizes on-the-fly patch validation to all existing state-of-the-art \apr systems, even including systems at the source code level.  This work also shows the first empirical evidence that on-the-fly patch validation can incur imprecise/unsound patch validation results, and further introduces a new technique for resetting JVM state for precise patch validation. The experimental results show that this work can speed up state-of-the-art representative \apr tools, including \capgen, \simfix, and \acs, by over an order of magnitude without incurring any imprecision.    
\bibliographystyle{ACM-Reference-Format}
\bibliography{bibdb}
\end{document}